\documentclass[12pt]{article}
\pdfoutput=1
\usepackage[pdftex]{graphicx}

\usepackage{amsmath}
\usepackage{mathtools}
\usepackage{amssymb}
\usepackage{physics}
\usepackage{braket}
\usepackage[subrefformat=parens]{subcaption}
\usepackage{float}
\numberwithin{equation}{section}
\usepackage{cite}
\usepackage[normalem]{ulem}
\usepackage{color}

\usepackage[
    colorlinks=true,
    linkcolor=blue,
    citecolor=blue,
    urlcolor=black
]{hyperref}

\begin{document}
\setlength{\baselineskip}{0.7cm}
\begin{titlepage}
\begin{flushright}
NITEP 274 \\
\end{flushright}
\vspace*{10mm}%
\begin{center}{\Large\bf
Two Higgs Doublet Model \\
\vspace*{2mm}
from Six-Dimensional Gauge Theory}
\end{center}
\vspace*{10mm}
\begin{center}
{\large Kento Akamatsu}$^{a}$,
{\large Takuya Hirose}$^{b}$, 
{\large Nobuhito Maru}$^{a,c}$ and
{\large Akio Nago}$^{a}$ 
\end{center}
\vspace*{0.2cm}
\begin{center}
${}^{a}${\it
Department of Physics, Osaka Metropolitan University, \\
Osaka 558-8585, Japan}
\\
${}^{b}${\it Faculty of Science and Engineering, Kyushu Sangyo University, \\
Fukuoka 813-8503, Japan}
\\
${}^{c}${\it Nambu Yoichiro Institute of Theoretical and Experimental Physics (NITEP), \\
Osaka Metropolitan University,
Osaka 558-8585, Japan}
\end{center}
\vspace*{1cm}

\begin{abstract}
We investigate a two-Higgs-doublet model (2HDM) based on a six-dimensional $SU(4)$ gauge theory compactified on $T^2/Z_2$.
The six-dimensional gauge symmetry ensures that the tree-level Higgs potential is CP conserving and $Z_2$ symmetric, 
while the Yukawa interactions of bulk fermions are naturally aligned, suppressing tree-level flavor-changing neutral currents. 
The $SU(4)$ charge embedding accommodates the quark electric charges, while the group-theoretical enhancement of the Yukawa coupling in the sixth-rank totally symmetric representation leads to the tree level relation $m_t\simeq 2m_W$.

We show that electroweak symmetry breaking can be realized and the observed SM Higgs boson mass can be reproduced through appropriate choices of the brane-localized gauge kinetic term coefficient and the bulk matter content. 
We also obtain predictions for the masses of the additional Higgs bosons. 
Furthermore, we show that the torus geometry affects both the Higgs mass spectrum and the couplings of the CP-even Higgs bosons to electroweak gauge bosons.
Finally, we identify parameter regions in which the additional Higgs mass spectra are consistent with the ATLAS and CMS measurements of the W-boson mass within the approximation based on the scalar-sector contribution to $\Delta \rho$.
\end{abstract}

\end{titlepage}

\newpage
\section{Introduction}

The discovery of the Higgs boson has completed the particle content of the Standard Model (SM), but the origin of the Higgs sector and the structure of its interactions remain open questions.
The two-Higgs-doublet model (2HDM) is one of the simplest extensions of the SM and predicts five physical Higgs bosons: two CP-even neutral scalars, one CP-odd neutral scalar, and a pair of charged scalars.
Its phenomenological implications have been extensively investigated, both theoretically and experimentally \cite{2HDMReview,2HDMexp,Wang:2022yhm}.
However, a general 2HDM requires additional assumptions to satisfy phenomenological constraints.
In particular, a discrete symmetry is commonly imposed on the Yukawa sector to suppress flavor-changing neutral currents (FCNCs) at tree level, while soft breaking of this symmetry is allowed in the scalar potential.
CP conservation is also frequently assumed to restrict the scalar interactions.
Furthermore, the conventional 2HDM does not provide a fundamental explanation for the hierarchy problem, and the Higgs masses and quartic couplings remain free parameters.

Gauge--Higgs unification (GHU) offers a different framework in which the Higgs fields originate from extra-dimensional
components of a higher-dimensional gauge field \cite{Manton,Fairlie,Hosotani1,Hosotani2,HIL,ABQ,MY,HMTY}.
In six-dimensional theories, two Higgs doublets can arise from the zero modes of the two extra-dimensional components of the gauge field, with their interactions constrained by the higher-dimensional gauge symmetry \cite{LMH,HLM,SSSW,CGM,MS,CKP,HNT,AHMN,SSS,KKKY,PT,BQ,KKY}. 
In these models, local tree-level mass terms for the Higgs doublets are forbidden by the higher-dimensional gauge symmetry, while the Higgs potential is generated through quantum corrections associated with Wilson-line phases.
The resulting Higgs mass is typically suppressed relative to the compactification scale by a loop factor.
Moreover, the effective potential is UV finite under appropriate conditions, providing a possible solution to the gauge hierarchy problem.
Since the compactification scale is also related to the electroweak gauge-boson masses, the Higgs and gauge sectors are closely connected.

In our previous study \cite{AHMN}, we constructed a 2HDM from a 6D $SU(4)$ gauge theory compactified on the orbifold $T^2/Z_2$ with square torus.
An important feature of this construction is that the tree-level Higgs potential is automatically CP conserving
and $Z_2$ symmetric, and its quartic couplings are fixed by the gauge interaction.
For an appropriate embedding of the electroweak generators, the model also predicts $\sin^2\theta_W=1/4$ at the compactification scale \cite{HLM}.
The Yukawa interactions of bulk fermions originate from the same six-dimensional gauge interaction, providing a natural realization of Yukawa alignment and thereby suppressing tree-level FCNCs.
Furthermore, the matrix elements of the gauge generators in the higher rank totally symmetric representation enhance the top-quark Yukawa coupling, leading to the tree-level relation $m_t\simeq 2m_W$.
Thus, the higher-dimensional gauge structure accounts for both the restricted Higgs potential and the aligned Yukawa interactions, which are commonly assumed in phenomenological 2HDMs.
Although electroweak symmetry breaking was successfully realized in our previous model, the predicted Higgs boson mass and  compactification scale were too low to satisfy experimental requirements.
To overcome this difficulty, we introduce brane-localized gauge kinetic terms (BLKTs) at the orbifold fixed points, which modify the KK spectrum and can raise the compactification scale \cite{SSS}.

In this paper, we extend our previous analysis of the six-dimensional $SU(4)$ gauge--Higgs unification model by incorporating BLKTs and generalizing the compactification geometry.
We investigate the resulting 2HDM phenomenology and compare the predicted Higgs spectrum with experimental constraints.
We show that the observed SM Higgs boson mass can be reproduced while enhancing the compactification scale to the TeV range through suitable choices of the BLKT coefficient and additional bulk scalar content.
Furthermore, we demonstrate that the torus geometry plays an important role in determining the couplings of the lighter CP-even Higgs boson to electroweak gauge bosons and the additional Higgs mass spectrum.
In particular, the lighter CP-even Higgs boson can acquire exactly SM-like gauge couplings as a consequence of the symmetry of the effective potential.
We also identify scalar mass spectra compatible with the ATLAS and CMS $W$-boson mass measurements within the scalar-sector $\Delta \rho$ approximation.

The remainder of this paper is organized as follows.
In Section 2, we introduce the six-dimensional model and derive its tree-level Higgs potential. 
In Section 3, we discuss the fermion representations and Yukawa interactions. 
In Section 4, we investigate the KK spectrum in the presence of brane-localized gauge kinetic terms.
In Section 5, we calculate the one-loop effective potential and study electroweak symmetry breaking and the Higgs spectrum.
In Section 6, we extend the analysis to a generalized torus and examine the alignment properties and electroweak precision constraints. 
The final section is devoted to our conclusions.

\section{Six-dimensional $SU(4)$ Gauge-Higgs Unification}

We consider a six-dimensional $SU(4)$ gauge theory compactified on the orbifold $T^2/Z_2$.
The six-dimensional coordinates are denoted by
	\begin{align}
		x^M=(x^\mu,x^5,x^6), \qquad M, N=0,1,2,3,5,6,
	\end{align}
where $x^\mu$ are the four-dimensional Minkowski coordinates and
$x^5$ and $x^6$ parametrize the two compact extra dimensions.
In this section, we first consider a square torus with equal compactification radii $R_1=R_2=R$,
	\begin{align}
		\begin{aligned}
			x^5 &\sim x^5+2\pi R, \\
			x^6 &\sim x^6+2\pi R .
		\end{aligned}
	\end{align}
The six-dimensional Lagrangian of gauge field is given by
	\begin{align}
		\mathcal{L}_{\rm gauge} = -\frac{1}{2} \Tr \qty( F_{MN}F^{MN}),
		\label{eq:gauge_lagrangian}
	\end{align}
where
	\begin{align}
		F_{MN} = \partial_M A^a_N - \partial_N A^a_M - i g_6 [A_M, A_N] ,
		\label{eq:field_strength}
	\end{align}
and $g_6$ denotes the six-dimensional gauge coupling constant. $a=1, \dots, 15$ labels the $SU(4)$ generators listed in Appendix~\ref{app:generator}.

\subsection{Orbifold boundary conditions}

The $T^2/Z_2$ orbifold is obtained by identifying
	\begin{align}
		(x^5,x^6)\sim(-x^5,-x^6).
	\end{align}
There are four fixed points,
	\begin{align}
		(x^5_i,x^6_i) = (0,0),\, (\pi R,0),\, (0,\pi R),\, (\pi R,\pi R).
		\label{eq:fixed_points}
	\end{align}
Around each fixed point $i = 0,1,2,3$, the $Z_2$ transformations for the gauge fields are defined by
	\begin{align}
		\begin{aligned}
			A_\mu(x^\mu,x^5_i-x^5,x^6_i-x^6) 
			&= P_i A_\mu(x^\mu, x^5_i+x^5,x^6_i+x^6) P_i^{-1} ,  \\
    			A_{5,6}(x^\mu,x^5_i-x^5,x^6_i-x^6)
    			&= -P_i A_{5,6}(x^\mu, x^5_i+x^5,x^6_i+x^6) P_i^{-1} .
		\end{aligned}  \label{eq:orbifold_bc_A56}
	\end{align}
Since the parity matrices satisfy
	\begin{align}
		P_3=P_1P_0P_2=P_2P_0P_1 ,
	\end{align}
we note that the parity matrix $P_3$ is not independent.
We choose the parity matrices as
	\begin{align}
		\begin{aligned}
			P_0 =& {\rm diag}(+1,+1,+1,-1),  \\
			P_1=P_2 &= {\rm diag}(+1,+1,-1,-1). 
		\end{aligned} \label{eq:P12}
	\end{align}
With this choice, the parity assignments of the gauge field components are given by
	\begin{align}
		\begin{aligned}
		A_\mu \to
			\begin{pmatrix}
 			(+++) & (+++) & (+--) & (---) \\
 			(+++) & (+++) & (+--) & (---) \\
 			(+--) & (+--) & (+++) & (-++) \\
 			(---) & (---) & (-++) & (+++)
			\end{pmatrix}, \\[3mm]
		A_{5,6} \to
			\begin{pmatrix}
			 (---) & (---) & (-++) & (+++) \\
			 (---) & (---) & (-++) & (+++) \\
			 (-++) & (-++) & (---) & (+--) \\
			 (+++) & (+++) & (+--) & (---)
			\end{pmatrix}.
		\end{aligned} \label{eq:parity_A56}
	\end{align}
Here the three signs denote the eigenvalues under $(P_0,P_1,P_2)$, while the parity at the fourth fixed point is determined by the consistency condition above.
Only the $(+++)$ components contain zero modes.
Therefore, the orbifold projection determines both the unbroken four-dimensional gauge symmetry and the scalar zero modes originating from the extra-dimensional components $A_5$ and $A_6$.

From the parity assignments in \eqref{eq:parity_A56}, the orbifold boundary conditions break the $SU(4)$ gauge symmetry as
\begin{align}
    SU(4) \xrightarrow{\text{$Z_2$-parity}} SU(2)_L \times U(1)_A \times U(1)_B
\end{align}
in the four-dimensional effective theory.
On the other hand, the zero modes of the extra-dimensional gauge fields are contained in the off-diagonal components,
\begin{align}
    A_{5,6}^{(0)}
    = \frac{1}{\sqrt{2}}
    \begin{pmatrix}
        0 & 0 & 0 & \phi^+_{1,2} \\
        0 & 0 & 0 & \phi^0_{1,2} \\
        0 & 0 & 0 & 0 \\
        \phi^-_{1,2} & {\phi^0}^*_{1,2} & 0 & 0
    \end{pmatrix}.
    \label{eq:A56_zero_modes}
\end{align}
These components transform as $SU(2)_L$ doublets and are identified with two Higgs doublets originating from $A_5$ and $A_6$, respectively. We parametrize the corresponding zero modes as
	\begin{align}
		\Phi_1 =
			\begin{pmatrix}
			\phi_1^+ \\ \phi_1^0
			\end{pmatrix}, \qquad
		\Phi_2 =
			\begin{pmatrix}
			\phi_2^+ \\ \phi_2^0
			\end{pmatrix}. \label{eq:higgs_doublets}
	\end{align}

After electroweak symmetry breaking, the neutral components develop vacuum expectation values,
	\begin{align}
		\ev{\Phi_1} = \frac{1}{\sqrt{2}}
			\begin{pmatrix}
			0 \\ v_1
			\end{pmatrix}, \qquad
		\ev{\Phi_2} = \frac{1}{\sqrt{2}}
			\begin{pmatrix}
			0 \\ v_2
			 \end{pmatrix}, \label{eq:higgs_vevs}
	\end{align}
with $v=\sqrt{v_1^2+v_2^2}=246 \ \text{GeV}$.

\subsection{Tree-level Higgs potential}

The tree-level Higgs potential originates from the extra-dimensional component of the six-dimensional gauge kinetic term.
For the zero modes, the derivatives with respect to the extra dimensions vanish, and hence
	\begin{align}
		F_{56}^{(0)}=-ig_6\ [A_5^{(0)},A_6^{(0)}].
	\end{align}
After dimensional reduction, the tree-level potential for the two Higgs doublets is given by
	\begin{align}
		V_{\rm tree}(\Phi_1,\Phi_2)
		=\frac{g_4^2}{2} \qty[ (\Phi_1^\dagger\Phi_1)(\Phi_2^\dagger\Phi_2)+(\Phi_1^\dagger\Phi_2)(\Phi_2^\dagger\Phi_1)-(\Phi_1^\dagger\Phi_2)^2 -(\Phi_2^\dagger\Phi_1)^2 ].
	\label{eq:tree_higgs_potential}
	\end{align}
The quartic interactions are fixed by the higher-dimensional gauge interaction. 
At tree level, the potential possesses a flat direction and contains no quadratic terms for the Higgs doublets.
The quadratic terms are generated radiatively through the Wilson-line effective potential and play an essential role in determining the electroweak vacuum.
Another important feature of Eq.~\eqref{eq:tree_higgs_potential} is its symmetry structure.
The tree-level potential is invariant under the discrete transformation
\begin{align}
    \Phi_1 \to \Phi_1,
    \qquad
    \Phi_2 \to -\Phi_2 ,
\label{eq:Higgs_Z2}
\end{align}
without assuming it by hand. 
In conventional 2HDMs, such a discrete symmetry is often introduced to control tree-level flavor-changing neutral currents (FCNCs) through its extension to the Yukawa sector. 
In the present model, the structure of the Yukawa interactions is instead determined by the higher-dimensional gauge interaction, as discussed in the next section.

The tree-level Higgs potential is also CP invariant. 
In a general 2HDM, complex parameters in the scalar potential can provide additional sources of CP violation, which are strongly constrained by electric dipole moment (EDM) measurements.
In the present model, the quartic interactions originate from the gauge kinetic term with the same gauge coupling constant.
Thus, the $Z_2$ and CP structures of the tree-level scalar sector are automatically preserved.
The $Z_2$ symmetry is softly broken by radiative corrections.
In particular, the Wilson-line effective potential generates the bilinear mixing term
	\begin{align}
		-m_{12}^2\left(\Phi_1^\dagger\Phi_2+\Phi_2^\dagger\Phi_1 \right),
	\label{eq:soft_Z2_breaking}
	\end{align}
while the tree-level quartic term remains constrained by its gauge symmetry. 
The resulting Higgs sector therefore realizes a CP-conserving 2HDM with a softly broken $Z_2$-symmetry without introducing these symmetries as independent assumptions.
The physical Higgs masses are determined by the curvature of the one-loop effective potential around the radiatively selected vacuum, as discussed in Sec. 5.
\footnote{
In six-dimensional gauge--Higgs unification, tadpole operators proportional to the field strength $F_{56}$ may arise at orbifold fixed points where a $U(1)$ symmetry is unbroken. 
If the tadpole term is allowed, it gives rise to the UV sensitive Higgs mass and we lose predictions for Higgs mass spectrum.
On the $T^2/Z_2$ orbifold, such operators can be forbidden by imposing an additional reflection symmetry under which $F_{56}$ is odd \cite{SSSW, CGM, BQ}.
We assume this symmetry and therefore do not include localized tadpole terms in our analysis.}

\section{Fermion sector and Yukawa interactions}
\label{sec:fermion}

In this section, we discuss the fermion sector and the origin of the Yukawa interactions. 
In gauge--Higgs unification, the Yukawa interactions originate from the higher-dimensional gauge interaction.
Since the two Higgs doublets arise from the extra-dimensional components $A_5$ and $A_6$ of the same gauge field, their Yukawa couplings are not independent.
In particular, the two Yukawa matrices in the up-type sector are proportional to each other due to the six-dimensional gauge interaction.
This leads to an alignment of Yukawa couplings, which prevents tree-level FCNCs in this sector.

\subsection{Electroweak charges and fermion embedding}
\label{subsec:fermion_embedding}

We first discuss the embedding of the electric charge into the $SU(4)$ gauge group. 
Since the Higgs doublets originate from the extra-dimensional components of the $SU(4)$ gauge field, they are
embedded in the adjoint representation.
Therefore, their electric charges are determined by the charge differences between the corresponding components of the fundamental representation.

Let the electric charge operator acting on the fundamental representation be
	\begin{align}
		Q_{\bf 4} = {\rm diag}(q_1,q_2,q_3,q_4).
	\label{eq:general_charge_4}
	\end{align}
The electric charge of an off-diagonal component $(A_M)^i{}_j$ of the adjoint representation is then given by 
	\begin{align}
		Q\!\left[(A_M)^i{}_j\right]=q_i-q_j .
	\label{eq:adjoint_charge}
	\end{align}
Requiring the zero modes of $A_5$ and $A_6$ identified in \eqref{eq:A56_zero_modes} to have the electric charges of Higgs doublet constrains the relative values of $q_i$. 
For instance, the $(i,j)=(2,4)$ or $(4,2)$ component of $A_5$ and $A_6$ must be electrically neutral, requiring $q_2-q_4=0$. 
Furthermore, since the electric charge generator is embedded in $SU(4)$, it must be traceless.
These conditions determine the electric charges of the fundamental representation up to one free parameter $q$ as
	\begin{align}
		Q_{\bf 4} = {\rm diag} \qty(q,\, q-1,\, 2-3q,\, q-1).
		\label{eq:charge_4}
	\end{align}
The charge assignment in Eq.~\eqref{eq:charge_4} also determines the embedding of the electroweak hypercharge.
The electric charge operator is defined as
	\begin{align}
	    Q_{\rm em}=I^3+\frac{Y}{2} ,
	\end{align}
where $I^3$ and $Y$ denote the third component of weak isospin and the weak hypercharge, respectively.
The hypercharge generator $Y$ is identified with a linear combination of the two Abelian generators associated with $U(1)_A$ and $U(1)_B$.
Its embedding into the bulk $SU(4)$ gauge group fixes the relative normalization of the $SU(2)_L$ and $U(1)_Y$ gauge couplings.
Consequently, the weak mixing angle at the compactification scale is predicted to be
	\begin{align}
		\sin^2\theta_W
		= \frac{{\rm{Tr}}\ I_3 ^2}{{\rm{Tr}}\ Q_{\bf 4}^2} 
		=\frac{1}{4(6q^2-8q+3)}.
		\label{eq:weak_mixing_angle}
	\end{align}
A phenomenologically viable embedding requires this tree-level prediction to be close to the observed weak mixing angle, $\sin^2\theta_W\simeq 0.23$. 
In particular, $\sin^2\theta_W=1/4$ is obtained for $q=1/3$ or $q=1$.
In this model, we adopt $q=1/3$
, which allows quarks with fractional electric charges to be embedded into $SU(4)$ representations.
The corresponding normalized diagonal generators of $SU(4)$ are
	\begin{align}
		\begin{aligned}
			T_L^3 &= \frac12 \mathrm{diag}(1,-1,0,0), \\[2mm]
			T_Y &= \frac{1}{6\sqrt{3}} \mathrm{diag}(-1,-1,6,-4), \\[2mm]
			T_X &= \frac{1}{6\sqrt{6}} \mathrm{diag}(5,5,-3,-7).
		\end{aligned}
	\end{align}
Here, $T_Y$ generates the electroweak hypercharge direction, while $T_X$ is the orthogonal Abelian generator.
The normalized electromagnetic generator in the fundamental representation is then given by
	\begin{align}
		T_{\rm em} &=\frac12 T_L^3+\frac{\sqrt{3}}{2}\,T_Y .
	\end{align}
For this choice of $q$, the electric charge operators in the fundamental and anti-fundamental representations are given by
	\begin{align}
		Q_{\mathbf{4}}=\mathrm{diag} \left(\frac13,-\frac23,1,-\frac23\right),\quad 
		Q_{\mathbf{4}^*}=-Q_{\mathbf{4}}.
	\label{eq:charge_fundamental}
	\end{align}
Thus, the anti-fundamental representation contains the electric charge assignments of the left-handed quark doublet and the right-handed up-type quark.

\subsection{Orbifold boundary conditions and chiral zero modes}
\label{subsec:fermion_bc}

The kinetic term for fermion in six-dimensional lagrangian is given by
	\begin{align}
		\mathcal{L}_{\rm fermion} = -\overline{\Psi}i\Gamma^M D_M\Psi, \qquad
		D_M=\partial_M-ig_6 A_M .\label{eq:fermion_6D}
	\end{align}
We adopt the following representation of the six-dimensional gamma matrices: 
	\begin{align}
		\Gamma^\mu =
			\begin{pmatrix}
			\gamma^\mu & 0\\
			0 & \gamma^\mu
			\end{pmatrix},\ 
		\Gamma^5 =
			\begin{pmatrix}
			 0 & i\gamma^5\\
			i\gamma^5 & 0
			\end{pmatrix},\ 
		\Gamma^6 =
			\begin{pmatrix}
			0& \gamma^5 \\
			-\gamma^5  & 0 
			\end{pmatrix}, \ 
		\gamma^5=
			\begin{pmatrix}
			0 & 1 \\
			1 & 0
			\end{pmatrix}.
	\label{eq:6D_gamma}
	\end{align}
We note that $\gamma^\mu,\gamma^5$ denote ordinary four-dimensional gamma matrices.
The six-dimensional chirality matrix is defined by
	\begin{align}
		\Gamma^7 \equiv \Gamma^0\Gamma^1\Gamma^2\Gamma^3\Gamma^5\Gamma^6 .
	\end{align}
With the convention in Eq.~\eqref{eq:6D_gamma}, it satisfies
	\begin{align}
		(\Gamma^7)^2=-1, \qquad \{\Gamma^7,\Gamma^M\}=0 .
	\end{align}
The projection operators onto the two six-dimensional chiralities are then defined as
	\begin{align}
		P_\pm = \frac{1}{2}\left(1\pm i\Gamma^7\right), \qquad
		\Psi_\pm=P_\pm\Psi , \label{eq:6D_chiral_projection}
	\end{align}
where $\Psi_+$ and $\Psi_-$ denote six-dimensional Weyl fermions with positive and negative chirality, respectively.
A six-dimensional Weyl fermion contains both left- and right-handed four-dimensional components:
	\begin{align}
		\Psi_+ = 
			\begin{pmatrix}
			\psi_L \\
			0 \\
			0 \\
			\psi_R
			\end{pmatrix}, \quad 
		\Psi_- = 
			\begin{pmatrix}
			0 \\
			\tilde{\psi}_R \\
			\tilde{\psi}_L \\
			0
			\end{pmatrix}.
	\end{align}

The chiral zero-mode spectrum in four dimensions is determined by the orbifold boundary conditions.
The orbifold boundary conditions for the bulk fermions are imposed consistently with those of the gauge fields.
Around each fixed point, we impose
	\begin{align}
		\Psi(x^\mu,x_i^5-x^5,x_i^6-x^6)
		&= \eta_i P_i(-i\Gamma^5\Gamma^6) \Psi(x^\mu,x_i^5+x^5,x_i^6+x^6) \notag \\[2mm]
		&=-\eta_i P_i \qty({\bf 1}_4\otimes\sigma_3)  \Psi(x^\mu,x_i^5+x^5,x_i^6+x^6) ,
		\label{eq:fermion_BC}
	\end{align}
where $\eta_i=\pm1$ denotes the intrinsic parity and $P_i$ acts on the corresponding $SU(4)$ representation.
The orbifold projection assigns opposite parities to the four-dimensional left- and right-handed components.
Using the parity matrices introduced in the gauge sector, the parity assignment for a fermion in the fundamental representation can be written as
	\begin{align}
		\Psi_{\bf 4}=
			\begin{pmatrix}
			\psi_{1L}^{(+++) }+\psi_{1R}^{(---)}\\[2mm]
			\psi_{2L}^{(+++) }+\psi_{2R}^{(---)}\\[2mm]
			\psi_{3L}^{(++-) }+\psi_{3R}^{(--+)}\\[2mm]
			\psi_{4L}^{(---) }+\psi_{4R}^{(+++) }
			\end{pmatrix}. \label{eq:fermion_parity_4}
	\end{align}
The orbifold projection leaves a left-handed $SU(2)_L$ doublet and a right-handed singlet in the four-dimensional effective theory.
With the charge assignment discussed above, these zero modes are identified with the left-handed quark doublet $Q_L$ and the right-handed up-type quark singlet $u_R$, respectively.
The resulting chiral zero modes constitute the bulk up-type quark sector, whose Yukawa interactions
are generated by the extra-dimensional gauge interaction.

\subsection{Yukawa interactions}
\label{subsec:gauge_yukawa}

A characteristic feature of gauge--Higgs unification is that the Yukawa interactions originate from the gauge interaction.
The extra-dimensional part of Eq.~\eqref{eq:fermion_6D} is
\begin{align}
\mathcal{L}_{\rm int} \supset g_6\overline{\Psi} \left( \Gamma^5 A_5+\Gamma^6 A_6 \right)\Psi .
\label{eq:yukawa_6D}
\end{align}
Since the zero modes of $A_5$ and $A_6$ are identified with $\Phi_1$ and $\Phi_2$, their
Yukawa interactions are not independent.
The gamma-matrix in \eqref{eq:yukawa_6D} selects the combination $A_6+iA_5$. 
Consequently, the two four-dimensional Yukawa matrices are proportional to each other,
\begin{align}
    Y_1 \propto Y_2 .
    \label{eq:general_yukawa_alignment}
\end{align}
This relation is known as Yukawa alignment \cite{PT}.
In our six-dimensional model, Yukawa alignment arises directly from the higher-dimensional gauge structure rather than being imposed as an additional assumption in the four-dimensional theory.

We first consider an anti-fundamental fermion $\Psi_{\mathbf{4}^*}$, whose orbifold projection gives the left-handed quark doublet $Q_L$ and the right-handed up-type quark singlet $u_R$ as chiral zero modes.
We define the charge-conjugated Higgs doublets as
	\begin{align}
		\widetilde{\Phi}_i \equiv i\sigma_2\Phi_i^*, \qquad i=1,2 ,
		\label{eq:tilde_phi}
	\end{align}
which have the appropriate quantum numbers for the up-type Yukawa interaction.
After dimensional reduction, the interaction among the zero modes is therefore given by
\begin{align}
\mathcal{L}_{Y}^{u}
&=-\overline{Q}_L\left( Y_1^u\widetilde{\Phi}_1 +Y_2^u\widetilde{\Phi}_2 \right)u_R +{\rm h.c.}\notag \\[2mm]
&= -\overline{Q}_L Y^u \left( i\widetilde{\Phi}_1 +\widetilde{\Phi}_2 \right) u_R +{\rm h.c.} ,
\label{eq:up_yukawa_GHU}
\end{align}
where $Y^u_{1,2}$ contain the group-theoretical factor and the overlap integral of the fermion zero-mode wave functions in the extra dimensions.
Thus, the two Yukawa matrices are automatically aligned in flavor space and can be simultaneously diagonalized, so that no tree-level FCNC arise from the bulk fermion sector.

While the two Yukawa interactions are fixed by the higher-dimensional gauge interaction, their overall magnitudes depend on the fermion representations and the localization profiles in the extra dimensions.
For fermions in the anti-fundamental representation, the Yukawa coupling is of the same order as the gauge
coupling and is not sufficiently large to reproduce the top quark mass.
We therefore embed the top quark sector into a higher-dimensional representation of $SU(4)$.

In particular, we employ the sixth-rank totally symmetric representation ${\bf{84}^*}$, for which the Yukawa coupling is enhanced by the corresponding group factor.
For the components identified with the top-quark zero modes, this enhancement gives
\begin{align}
m_t \simeq 2m_W , \label{eq:top_mass}
\end{align}
at tree level. 
The explicit decomposition of the ${\bf 84^*}$ representation, the relevant generator matrix elements 
and the derivation of the above enhancement factor $2$ are given in Appendix~\ref{app:fermion_representation}.

The first- and second-generation up-type quarks are instead embedded in anti-fundamental representations.
To account for their small Yukawa couplings, we introduce bulk mass terms that localize the corresponding fermion zero modes along the fifth dimension:
\begin{align}
{\cal L}^{6D}_{f}= \bar{\Psi}_{f}\ i\Gamma^{M}D_{M}\Psi_{f}- \bar{\Psi}_{+}M_f \epsilon(x^5)\Gamma^6 \Psi_{+},
\qquad
f=u,c .
\label{6DkinkLagrangian}
\end{align}
Here, $\Psi_f$ denotes a six-dimensional Dirac fermion, while $\Psi_+$ is its positive-chirality Weyl component.
The parameter $M_f$ controls the localization of the fermion zero modes, and $\epsilon(x^5)$ denotes a periodic sign function along the fifth dimension.
Under the orbifold reflection, both $\epsilon(x^5)$ and $\bar{\Psi}_{+}\Gamma^6\Psi_{+}$ are odd.
Their product is therefore even, so the bulk mass term is compatible with the orbifold boundary conditions.
The resulting mass profile can localize the left- and right-handed fermion zero modes in different regions of the compact space.
Solving the Dirac equations, the zero mode functions for $M_f>0$ are given by
\begin{align}
\begin{aligned}
&f_{f {R}}^{(0)}(x^5) = \sqrt{\frac{2M_f}{1-e^{-2\pi M_f R_1}}}\  e^{-M_f x^5}, \\[3mm]
&f_{f {L}}^{(0)}(x^5) = \sqrt{\frac{2M_f}{1-e^{-2\pi M_f R_1}}}\  e^{-M_f (\pi R_1 - x^5)}. 
\end{aligned}
\end{align}
Since the Yukawa interactions originate from the higher-dimensional gauge interaction, their effective four-dimensional couplings are determined by the overlap of the corresponding zero-mode wave functions.
For sufficiently large bulk masses, the effective four-dimensional Yukawa couplings are exponentially suppressed, yielding
\begin{align}
    Y_f^{\rm eff} &= 
    N_f^2 \int_0^{\pi R_1} dx^5 f_{fL}^{(0)}(x^5) f_{fR}^{(0)}(x^5) \notag \\[2mm]
     &\propto 2\pi M_f R_1 \ e^{-\pi R M_f},
    \qquad f=u,c ,
\end{align}
where $N_f$ is the normalization constant of wave function.
In fact, if we choose the bulk masses as $M_u R_1 \sim 15, M_c R_1 \sim 8$, 
the up- and charm-quark masses can be reproduced by suitably choosing the bulk masses.
This mechanism provides the hierarchy of the up-type quark masses while preserving the gauge origin of the Yukawa interactions and the suppression of tree-level FCNCs in the up-sector.

{As for the down-type quark and the charged lepton Yukawa couplings, 
they are introduced on the fixed point ($(x^5, x^6)=(0, 0)$ for example) as 4D localized Yukawa couplings. 
\begin{align}
\label{localized_SM_fermion}
\mathcal{L}_{Y}^{d,l}(4D)
&= -\overline{Q}_L(0, 0) Y^d \left( i\widetilde{\Phi}_1(0,0) +\widetilde{\Phi}_2(0,0) \right) d_R +{\rm h.c.} \notag \\[2mm]
&\hspace*{2em}-\overline{L}_L Y^l \left( i\widetilde{\Phi}_1(0,0) +\widetilde{\Phi}_2(0,0) \right) e_R +{\rm h.c.} ,
\end{align}
where the right-handed down-type quarks and the charged leptons are introduced as four-dimensional fields rather than six-dimensional fields. 
The notaion $Q_L(0, 0)$, $\tilde{\Phi}_{1,2}(0,0)$denotes the values of the corresponding six-dimensional fields $Q_L(x^\mu, x^5, x^6)$, $\tilde{\Phi}_{1,2}(x^\mu, x^5,x^6)$ at $(x^5, x^6)=(0, 0)$, respectively. 
$Y^{d, l}_{1,2}$ are Yukawa couplings for the down-type quark and the charged leptons, respectively. 
Although such Yukawa coupling constants are arbitrary in principle, $Y_1^{d, l}=Y_2^{d, l}=Y^{d,l}$ is assumed in the \eqref{localized_SM_fermion} to avoid tree-level FCNC.  
Since these Yukawa couplings are essentially similar to those in the SM, 
the mass hierarchy for down-type quarks and the charged leptons is obtained by tuning the Yukawa coupling constants as in the SM. }

\section{Brane-localized gauge kinetic terms}
\label{sec:BLKT}

In orbifold theories, translational invariance in the extra dimensions is explicitly broken at the fixed points.
Consequently, operators localized at the fixed points are generically allowed, provided that they respect the gauge symmetry preserved locally at each fixed point \cite{GL,AVS1,AVS2,AC}.
In particular, gauge kinetic terms localized at the orbifold fixed points can appear in the six-dimensional
effective theory.
Such localized operators are also relevant from the viewpoint of renormalization.
In six-dimensional orbifold gauge theories, radiative corrections can generate divergences localized at the fixed points, which require localized gauge kinetic counterterms.
In addition, higher-derivative bulk counterterms 
can be generated. 
However, their effects are suppressed in the low-energy regime relevant to the electroweak gauge bosons.
Therefore, we focus on the brane-localized gauge kinetic terms (BLKTs), which directly modify the normalization of the gauge fields and their KK spectrum.

The finite parts of the coefficients of these localized terms are sensitive to the UV completion and are not determined within the low-energy effective theory.
Therefore, we treat them as effective parameters of the orbifold theory.
The Lagrangian for gauge fields including the BLKT is given by
	\begin{align}
		\mathcal{L}_{\rm gauge} 
		=&-\frac{1}{2} {\rm Tr}\left(F_{MN}F^{MN}\right) \notag \\
		&\quad -\frac{1}{4} \sum_i 2\pi^2 R^2 c_i\, \delta(x^5-x_i^5) \delta(x^6-x_i^6) F^{a_i}_{\mu\nu}F^{a_i \mu\nu},
	\label{eq:BLKT_action}
	\end{align}
where $c_i$ are dimensionless coefficients associated with the four orbifold fixed points $(x_i^5,x_i^6)$.

We next consider the effect of the BLKTs on the KK spectrum.
In the presence of a Wilson-line background, we define the dimensionless KK eigenvalues in the absence of the BLKTs as
	\begin{align}
		&E_{l_1,l_2}^{2} \equiv 
		\left( l_1+\frac{q\alpha_1+\varepsilon}{2} \right)^2 + \left( l_2+\frac{q\alpha_2+\varepsilon}{2} \right)^2
	\label{eq:KK_eigenvalue}, \ (l_1,l_2 = 0,\ \pm 1, \ \pm 2, \ \dots).
	\end{align}
Here, $\alpha_i=g_4v_iR$ ($i=1,2$) denote the dimensionless Wilson-line phases, where $g_4$ is the four-dimensional gauge coupling and $v_i$ are the corresponding vacuum expectation values.
Here, $q$ denotes the Wilson-line charge.
The parameter $\varepsilon$ is determined by the boundary condition under translation, which follows from the orbifold parity assignments.
It takes $\varepsilon=0$ for periodic modes and $\varepsilon=1$ for antiperiodic modes.
In the presence of the BLKTs, the KK masses are determined by
	\begin{align}
		\det \qty[ \delta_{ij}-M^2 c_i\ {\cal{G}}_{ji}(M) ]=0,
	\label{eq:BLKT_mass_condition}
	\end{align}
where ${\cal{G}}_{ji}(M)$ is the Green function evaluated between the two fixed points $i$ and $j$.
It is defined by
	\begin{align}
		{\cal{G}}_{ji}(M) \equiv R^2 \sum_{l_1,l_2} \frac{ \exp \left[ i\left(l_1+\frac{q\alpha_1+\varepsilon}{2}\right) \frac{x_j^5-x_i^5}{R} +i\left(l_2+\frac{q\alpha_2+\varepsilon}{2}\right) \frac{x_j^6-x_i^6}{R}\right]}{ E_{l_1,l_2}^2-M^2R^2}.
	\label{eq:BLKT_Green}
	\end{align}
The derivation of Eq.~\eqref{eq:BLKT_mass_condition} from the bulk equation of motion and the boundary-localized terms is given
in Appendix~\ref{app:BLKT_equation}.
To determine the lightest mode, we expand the Green function around the corresponding pole in Eq.~\eqref{eq:BLKT_mass_condition}.
Denoting the corresponding dimensionless squared eigenvalue in the absence of the BLKTs by $E^2_{0,0}$,
the leading pole contribution gives
	\begin{align}
		M^2_{\rm light} \simeq \frac{E_{0,0}^2}{(1+c) R^2},
		\qquad c\equiv\sum_{i=0}^{3}c_i . \label{eq:BLKT_light_mode}
	\end{align}
Thus, at leading order, the shift of the lightest mode depends only on the sum of the BLKT coefficients, rather than on their individual
distribution among the four fixed points.
In contrast, the excited KK spectrum generally depends on the individual $c_i$ through the propagation between different fixed points.
Motivated by this observation,  we consider a single BLKT localized at the fixed point $(x^5,x^6)=(0,0)$ in the following analysis,
	\begin{align}
		c_0=c, \qquad c_1=c_2=c_3=0.
	\label{eq:single_BLKT}
	\end{align}
This choice provides a simple parametrization of the leading low-energy effect and does not represent an assumption about the
UV origin of the BLKT.
With the choice in \eqref{eq:single_BLKT}, the KK mass condition reduces to
	\begin{align}
		1-c \xi^2 \ \sum_{l_1,l_2} \frac{1}{E^2_{l_1,l_2}-\xi^2} =0, \qquad \xi \equiv MR.
	\label{eq:single_BLKT_KK}
	\end{align}
To investigate the effects of the BLKTs on the KK spectrum, we numerically solve the mass equation \eqref{eq:single_BLKT_KK} for several values of the BLKT coefficient $c$.
In the numerical analysis, we regularize the logarithmically divergent two-dimensional lattice sums and retain their finite parts.
The resulting dimensionless KK eigenvalues are compared in Table~\ref{tab:KK_eigenvalues} for $(\alpha_1,\alpha_2)=(0.2,0.1)$ and $(q,\varepsilon)=(1,0)$.
As shown in Table~\ref{tab:KK_eigenvalues}, the lowest KK mass is substantially reduced by the BLKT and approximately scales as $1/\sqrt{1+c}$.
In contrast, several excited KK modes exhibit much smaller shifts, indicating that the BLKT does not uniformly rescale the KK spectrum.
The BLKT also modifies the normalization of the effective four-dimensional gauge coupling.
The relation between the six- and four-dimensional gauge couplings is given by
	\begin{align}
		g^2_6 \simeq 2 \pi^2 R^2 (1+c)\,g^2_4 .
	\label{eq:g6_g4_BLKT}
	\end{align}
Thus, the BLKT modifies not only the massive KK spectrum but also the normalization of the four-dimensional gauge interaction.
For the $W$ boson, corresponding to $(q,\varepsilon)=(1,0)$, Eq.~\eqref{eq:BLKT_light_mode} gives
	\begin{align}
		m_W^2 \simeq \frac{\alpha_1^2+\alpha_2^2}{4(1+c)R^2}.
	\label{eq:MW_BLKT}
	\end{align}
This approximation is valid for sufficiently small Wilson-line phases.
In our numerical analysis, however,  the $W$-boson mass is determined from the lowest solution of the mass equation \eqref{eq:single_BLKT_KK}, without relying on this approximation.
\begin{table}[http]
\centering
\caption{
Comparison of the dimensionless KK eigenvalues with BLKTs for $(\alpha_1,\alpha_2)=(0.2,0.1)$, $(q,\varepsilon)=(1,0)$.
Here, $E_{l_1,l_2}$ denotes the BLKT-free spectrum, while $\xi_n$ denotes the KK eigenvalues including BLKT effect.}
\label{tab:KK_eigenvalues}
\begin{tabular}{cc|cc|cc|cc}
& & \multicolumn{2}{c|}{$c=0.1$} & \multicolumn{2}{c|}{$c=3$} & \multicolumn{2}{c}{$c=15$} \\
\hline
 & & & & & \\[-11pt]
$(l_1,l_2)$ & $E_{l_1,l_2}$ & $\xi_n$ & $\xi_n/E_{l_1,l_2}$ & $\xi_n$ & $\xi_n/E_{l_1,l_2}$& $\xi_n$ & $\xi_n/E_{l_1,l_2}$\\
 & & & & & \\[-11pt]
\hline
 & & & & & \\[-11pt]
$(0,0)$ & 0.11180 & 0.10656 & 0.95311 & 0.05455 & 0.48789  & 0.02693 & 0.24084   \\
$(-1,0)$ & 0.90139 & 0.79399 & 0.88085 & 0.41330 & 0.45852 & 0.37994 & 0.42151 \\
$(0,-1)$ & 0.95525 & 0.92890 & 0.97242 & 0.92203 & 0.96522 & 0.92186 & 0.96504 \\ 
$(0,1)$ & 1.05475 & 1.01431 & 0.96165 & 0.99665 & 0.94491  & 0.99617 & 0.94446 \\
 $(1,0)$ &1.10114 & 1.08286 & 0.98340 & 1.07847 & 0.97942 & 1.07834 & 0.97930 \\
$(-1,-1)$ & 1.30863 & 1.24574 & 0.95194 & 1.20377 & 0.91988 & 1.20245 & 0.91887 \\
$(-1,1)$ & 1.38293 & 1.34992 & 0.97613 & 1.34127 & 0.96987 & 1.34103 & 0.96970 \\
$(1,-1)$ & 1.45344 & 1.42644 & 0.98142 & 1.41952 & 0.97666 & 1.41932 & 0.97652 \\
\vdots &\vdots &\vdots &\vdots  &\vdots &\vdots &\vdots &\vdots  
\end{tabular}
\end{table}

\section{One-loop effective potential and Higgs spectrum}
\label{sec:effective_potential}
In this section, we investigate electroweak symmetry breaking and the Higgs mass spectrum in the six-dimensional $SU(4)$ gauge--Higgs unification model.
We first evaluate the one-loop effective potential on the square torus in the absence of the brane-localized gauge kinetic terms.
We then determine the electroweak vacuum and examine the effects of the BLKTs on the vacuum structure.
Finally, we calculate the physical Higgs mass spectrum, taking into account the effect of BLKT on the gauge coupling normalization and the compactification scale.

\subsection{One-loop effective potential}
\label{subsec:one_loop_potential}

We evaluate the one-loop effective potential for the Wilson-line phases arising from the $SU(4)$ gauge field
and a bulk fermion in the sixth-rank totally symmetric representation $\mathbf{84}^*$.
In this subsection, we first consider the case without the BLKTs.
The one-loop contribution is given by the Coleman--Weinberg potential,
	\begin{align}
		V_{\rm 1-loop}
		= \frac{1}{2}\sum_I(-1)^{F_I}n_I \sum_{n_1,n_2} \int\frac{d^4p_E}{(2\pi)^4}  \log  \left[ p_E^2+M_{n_1,n_2}^{2} \right],
	\label{eq:CW_potential}
	\end{align}
where $F_I$ denotes the fermion number and $n_I$ counts the physical degrees of freedom.

A typical KK spectrum is given by
	\begin{align}
		M_{n_1,n_2}^2= \frac{1}{R^2} \left[ \left(n_1+\frac{q\alpha_1+\varepsilon}{2}\right)^2 + \left(n_2+\frac{q\alpha_2+\varepsilon}{2}\right)^2 \right].
	\end{align}
For periodic mode ($\varepsilon=0$), Poisson resummation yields the following contribution to the effective potential:
	\begin{align}
		V^{(I)}_{\rm 1-loop}(\alpha_1,\alpha_2)
		= -\frac{(-1)^F n_I}{16\pi^7R^4}
		 \sum_{(k_1,k_2)\neq(0,0)} \frac{ \cos(q\pi (k_1\alpha_1+ k_2\alpha_2))}{(k_1^2+k_2^2)^3}.
	\label{eq:CW_winding}
	\end{align}
Here, the Wilson-line-independent zero-winding contribution has been omitted.
For convenience,  we introduce the following function:
\begin{align}
\nu_{k_1,k_2}(\alpha_1,\alpha_2; q)
\equiv \frac{ \cos(q\pi (k_1\alpha_1+ k_2\alpha_2))}{(k_1^2+k_2^2)^3}.
\label{eq:winding_function}
\end{align}
The boundary conditions of the bulk fields determine the Wilson-line charges and the shifts of the KK momenta.
For antiperiodic modes ($\varepsilon=1$), the corresponding KK momentum is shifted according to
$q\alpha_i\ \to \  q\alpha_i+1$.
This shift introduces an additional factor $(-1)^{k_1+k_2}$ into the function.

For the $SU(4)$ gauge field, the adjoint representation decomposes into $SU(2)$ representation as 
\begin{align}
\begin{aligned}
{\bf{15}} \ \xrightarrow{SU(2)}\ 
&{\bf{3}}_{(+++)}\oplus {\bf{2}}_{(+--)}\times 2\oplus{\bf{2}}_{(---)}\times 2 \oplus{\bf{1}}_{(+++)} \times 2 \oplus {\bf{1}}_{(--+)}\times 2
\end{aligned}
\end{align}
Here, the subscripts denote the orbifold parities $(P_0,P_1,P_2)$.
Taking into account the orbifold boundary conditions and the four physical degrees of freedom of a six-dimensional gauge field, the gauge contribution to the one-loop effective potential is given by
\begin{align}
&V^{(g)}(\alpha_1,\alpha_2) \notag \\
& \quad = -\frac{1}{4\pi^7R^4} \sum_{(k_1,k_2)\neq(0,0)} \bigg[ \nu_{k_1,k_2}(\alpha_1,\alpha_2;2) +2\left(1+(-1)^{k_1+k_2}\right) \nu_{k_1,k_2}(\alpha_1,\alpha_2;1) \bigg].
\label{eq:gauge_potential}
\end{align}

We next consider the fermionic contribution to the effective potential.
As discussed in Sec.~\ref{sec:fermion}, the top quark is embedded in a bulk fermion transforming in the sixth-rank totally symmetric representation $\mathbf{84}^*$ of $SU(4)$.
Under $SU(2)_L$, this representation decomposes as
\begin{align}
\begin{aligned}
\mathbf{84^*}\ \xrightarrow{SU(2)} \;
& \mathbf{7}_{(+++)} \oplus \mathbf{6}_{(---)} \oplus \mathbf{6}_{(+--)} \oplus \mathbf{5}_{(+++)}\times 2 \oplus \mathbf{5}_{(-++)}\\
& \oplus \mathbf{4}_{(---)} \times 2 \oplus \mathbf{4}_{(+--)}\times 2 \oplus \mathbf{3}_{(+++)} \times 3 +\mathbf{3}_{(-++)} \times 2 \\
&\quad \oplus \mathbf{2}_{(---)}\times 3 \oplus \mathbf{2}_{(+--)} \times 3 \oplus \mathbf{1}_{(+++)}\times 4 \oplus \mathbf{1}_{(-++)}\times 3.
\end{aligned}
\end{align}
Taking into account the eight degrees of freedom of a six-dimensional Dirac fermion and the color multiplicity $N_c=3$, its contribution to the one-loop effective potential is given by
\begin{align}
&V^{(f_{84^*})}(\alpha_1,\alpha_2) \notag \\
&\quad = \frac{3}{2\pi^7R^4} \sum_{(k_1,k_2)\neq(0,0)} 
\bigg[ \nu_{k_1,k_2}(\alpha_1,\alpha_2;6) +\left(1+(-1)^{k_1+k_2}\right) \nu_{k_1,k_2}(\alpha_1,\alpha_2;5) \notag \\[1mm]
&\quad  \hspace{1cm} +\left(3+(-1)^{k_1+k_2}\right) \nu_{k_1,k_2}(\alpha_1,\alpha_2;4)+3\left(1+(-1)^{k_1+k_2}\right) \nu_{k_1,k_2}(\alpha_1,\alpha_2;3) \nonumber\\
&\ \hspace{1.5cm} +3\left(2+(-1)^{k_1+k_2}\right) \nu_{k_1,k_2}(\alpha_1,\alpha_2;2)+6\left(1+(-1)^{k_1+k_2}\right) \nu_{k_1,k_2}(\alpha_1,\alpha_2;1)
\bigg].
\label{eq:fermion84_potential}
\end{align}
In contrast, the first two generations of up-type quarks are embedded into the antifundamental representation and possess bulk masses.
For a bulk fermion with mass $M_f$, its contribution to the Wilson-line-dependent effective potential is exponentially suppressed by factors of the form $e^{-2\pi M_f R |\boldsymbol{k}|}$, where $\boldsymbol{k}=(k_1,k_2)$ denotes the winding numbers.
For large bulk masses, the contributions from the two generations are negligible compared with the contribution from the $\mathbf{84}^*$ fermion .
The down-type quarks and charged leptons acquire their masses through localized Yukawa interactions, as discussed in Sec.~\ref{sec:fermion}. Since their Yukawa couplings are much smaller than that of the top quark, we neglect their contributions to the one-loop effective potential in the present analysis.

Therefore, the total one-loop effective potential is given by
\begin{align}
V_{\rm 1-loop}(\alpha_1,\alpha_2)
= V^{(g)}_{\rm 1-loop}(\alpha_1,\alpha_2) + V^{(f_{84^*})}_{\rm 1-loop}(\alpha_1,\alpha_2).
\label{eq:total_one_loop_potential}
\end{align}
For the square torus, the effective potential is invariant under the shifts $\alpha_i \to \alpha_i +2$ and $\alpha_1 \leftrightarrow \alpha_2$, as well as the independent reflections $\alpha_i\to-\alpha_i$ ($i=1,2$).
These symmetries allow us to restrict the Wilson-line phases to the fundamental region
\begin{align}
0\leq\alpha_2 \leq \alpha_1  \leq 1.
\end{align}

\subsection{Electroweak symmetry breaking}
\label{subsec:EWSB}

In gauge--Higgs unification, electroweak symmetry breaking is induced by the Hosotani mechanism, in which the
Wilson-line phases acquire nonvanishing vacuum expectation values through quantum corrections.

\begin{figure}[http]
		\begin{tabular}{cc}
		\begin{minipage}{.5 \textwidth}
				\centering
				\includegraphics[width=8cm]{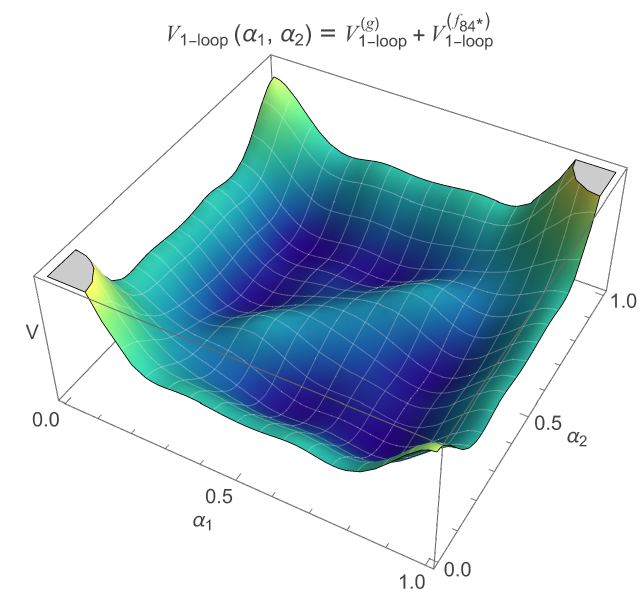}
			\end{minipage}
		\begin{minipage}{.5 \textwidth}
				\centering
				\includegraphics[width=6.5cm]{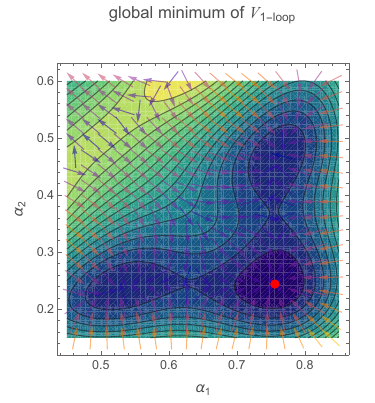}
			\end{minipage}
		\end{tabular}
		\caption{One-loop effective potential for the Wilson-line phases, including contributions from
the $SU(4)$ gauge field and the bulk fermion in the $\mathbf{84}^*$ representation. 
The left panel shows the 3D potential, while the right panel shows the contour plot around the global minimum.}
		\label{fig:potential84rep}
	\end{figure}

The vacuum configuration is determined by minimizing the one-loop effective potential obtained in the previous
subsection,
\begin{align}
\eval{\frac{\partial V_{\rm 1-loop}}{\partial\alpha_i} }_{(\alpha_1^{\rm min},\alpha_2^{\rm min})}=0.
\label{eq:vacuum_condition}
\end{align}
The global minimum specifies the vacuum expectation values of the Wilson-line phases and determines the pattern of electroweak symmetry breaking. Figure~\ref{fig:potential84rep} shows the one-loop effective potential as a function of the Wilson-line phases $\alpha_1$ and $\alpha_2$.
Although several local minima exist, a representative global minimum within the fundamental region is found at $(\alpha_1^{\min},\alpha_2^{\min})\simeq(0.756,0.244)$.

The unbroken gauge symmetry is determined by the generators that commute with the vacuum expectation values of the
Wilson-line operators.
For the two compact directions, we define
\begin{align}
\ev{W_{1,2}} 
&= \mathcal{P} \exp \left( ig_6\oint_{S^1}dx^{5,6}\,\ev{A_{5,6}} \right) \notag \\[2mm]
&=
\begin{pmatrix}
1 & 0 & 0 & 0\\
0 & \cos(\pi\alpha_{1,2}^{\rm min})
  & 0 & i\sin(\pi\alpha_{1,2}^{\rm min})\\
0 & 0 & 1 & 0\\
0 & i\sin(\pi\alpha_{1,2}^{\rm min})
  & 0 & \cos(\pi\alpha_{1,2}^{\rm min})
\end{pmatrix},
\label{eq:Wilson_line_vacuum}
\end{align}
where $S^1$ denotes the corresponding cycle of the torus.

A generator $\mathcal{T}$ of the unbroken gauge symmetry must preserve the orbifold boundary conditions and satisfy
\begin{align}
[\mathcal{T},\ev{W_1}]
=
[\mathcal{T},\ev{W_2}]
=0.
\label{eq:unbroken_generator_condition}
\end{align}
We find that two independent generators commute with the Wilson-line operators: the electromagnetic generator
$T_{\rm em}$ and another generator given by
\begin{align}
T^\prime
&=\frac{1}{6\sqrt{2}} \left(6T_L^3-2T_Y+\frac13 T_X\right) \notag \\[2mm]
&=\frac{1}{6\sqrt{2}} \mathrm{diag}(5,-1,-3,-1).
\end{align}
Thus, the Wilson-line vacuum leaves two Abelian gauge symmetries unbroken:
\begin{align}
SU(4)
\ \xrightarrow[\hspace{1.2cm}]{T^2/Z_2} \ 
SU(2)_L \times U(1)_Y \times U(1)_X
\ \xrightarrow[\hspace{1.2cm}]{\ev{W_{1,2}}} \ 
U(1)_{\rm em} \times U(1)^\prime .
\end{align}
The remaining $U(1)'$ gauge symmetry must be broken to avoid an additional massless gauge boson.
This can be achieved, for example, by introducing a suitable brane-localized scalar field whose vacuum expectation value breaks $U(1)'$.

\subsection{Effects of BLKT}
We next examine the effects of the BLKT on the minimum point  and the curvature of the one-loop effective potential.
As discussed in Sec.~\ref{sec:BLKT}, the BLKTs modify the gauge-field KK spectrum and hence its contribution to the one-loop effective potential.
Since Poisson resummation cannot be directly applied, we evaluate the dimensionless effective potential numerically using the KK masses obtained from the BLKT mass equation.
Introducing the dimensionless cutoff $s=\Lambda R$, the one-loop effective potential can be written as
\begin{align}
{\mathcal{V}}_{\rm 1-loop}(\alpha_1,\alpha_2; c) = -\frac{1}{32\pi^2} \sum_I (-1)^{F_I}n_I \sum_n \mathcal{I}\left(\xi_{I,n}(c);s\right),
\label{eq:BLKT_effective_potential}
\end{align}
where $\xi_{I,n}(c)$ denotes the dimensionless KK mass obtained by solving the mass equation and
\begin{align}
\mathcal{I}(\xi;s)
&\equiv \int_0^{s^2}dl\,l \exp\left(-\frac{\xi^2}{l}\right) \notag \\[2mm]
&=\xi^4 \ \Gamma\left(-2,\frac{\xi^2}{s^2}\right).
\label{eq:BLKT_integral}
\end{align}
Here, $\Gamma(a,z)$ is the upper incomplete gamma function.
The BLKT is introduced only for the gauge fields, while the fermionic contribution remains unchanged.
We therefore evaluate the BLKT correction to the one-loop effective  potential as
\begin{align}
\Delta\mathcal V^{}_{} = -\frac{1}{8 \pi^2} \sum_{I \in g}n_I \sum_n \qty[ \mathcal{I}\left(\xi_{I,n}(c);s\right)-\mathcal{I}\left(\xi_{I,n}(0);s\right)].
\end{align}

We compare the effective potentials obtained with and without the BLKTs to quantify their effects on the global minimum and curvature.
We define the curvature matrix as
\begin{align}
\mathcal H_{ij}(c)
\equiv 
\left.
\frac{\partial^2\mathcal{V}_{\rm 1-loop}(\alpha_1,\alpha_2;c)}{\partial\alpha_i\partial\alpha_j}
 \right|_{(\alpha_1^{\rm min},\alpha_2^{\rm min})}.
\label{eq:BLKT_hessian}
\end{align}

In the numerical evaluation, we set the dimensionless UV cutoff to $s^2=10$ and truncate the KK-mode sum to states satisfying $\xi_{I,n} < 15$.
As shown in Table~\ref{tab:BLKT_effective_potential},  the global minimum point remains nearly unchanged even if the BLKT coefficient increases.
For $c=50$, the minimum shifts by only $\Delta\alpha_i^{\rm min}\simeq 1.5\times10^{-3}$.
The smaller eigenvalue of the curvature matrix decreases by approximately $3.0\%$, corresponding to an estimated mass shift of about $1.9$ GeV for SM-like Higgs boson.
Such a correction may be compensated by readjusting the BLKT coefficient $c$, although the required adjustment depends on the full mass relation.
These results indicate that the BLKTs have only a small effect on the vacuum configuration and induce moderate corrections to the curvature of the effective potential.

In the following analysis of the scalar mass spectrum and the compactification scale, we therefore adopt the BLKT-free effective potential as a leading-order approximation, while retaining the BLKT effects on the
gauge KK spectrum and the four-dimensional effective couplings.
For more precise mass predictions, the BLKT-induced corrections to the curvature should be included, together
with a corresponding readjustment of the BLKT coefficient.
\begin{table*}[http]
\centering
\caption{Effects of the BLKT on the minimum point and the curvature.
$\mathcal{V}_0$ denotes the minimum value of the effective potential at $c=0$.
$\lambda_1$ and $\lambda_2$ are the eigenvalues obtained from the curvature matrix $\mathcal{H}_{ij}$.}
\label{tab:BLKT_effective_potential}
\begin{tabular}{ccccc}
\hline
&&&& \\[-11pt]
$c$
& $(\alpha_1^{\rm min},\alpha_2^{\rm min})$
& $\Delta \mathcal{V}_{\rm min} /\mathcal{V}_{0}$  [\%]
& $(\lambda_1,\lambda_2)$
& $(\Delta\lambda_1, \Delta\lambda_2)$ \\
&&&& \\[-11pt]
\hline
$0$
& $(0.75591,0.24409)$
& $0$
& $(0.28253,0.35929)$
& $(0,0)$
\\
$0.1$
& $(0.75556,0.24444)$
& $-0.000789$
& $(0.28055,0.35740)$
& $(-0.001973,-0.001884)$
\\
$3$
& $(0.75455,0.24545)$
& $-0.011902$
& $(0.27485,0.35216)$
&  $(-0.007673,-0.007125)$
\\
$10$
& $(0.75443,0.24557)$
& $-0.014012$
& $(0.27420,0.35160)$
&  $(-0.008326,-0.007687)$
\\
$50$
& $(0.75438,0.24562)$
& $-0.014917$
& $(0.27394,0.35138)$
&  $(-0.008590,-0.007911)$
\\
\hline
\end{tabular}
\end{table*}

\subsection{Higgs mass spectrum}
\label{subsec:higgs_spectrum}

We now investigate the masses of the physical Higgs bosons arising from the two Higgs doublets $\Phi_1$ and $\Phi_2$.
The effective potential is $V_{\rm eff}=V_{\rm tree}+V_{\rm 1-loop}$.
As discussed in Sec.~2, the tree-level Higgs potential contains only quartic interactions, whose couplings are fixed by the higher-dimensional gauge symmetry.
The quadratic terms are generated radiatively through the one-loop effective potential, thereby triggering electroweak symmetry breaking.
Since the tree-level potential vanishes identically along the neutral CP-even directions, the CP-even Higgs mass matrix is determined by the curvature of the one-loop effective potential.
Expanding the two Higgs doublets around their vacuum expectation values,
\begin{align}
\Phi_i=
\begin{pmatrix}
\phi_i^+\\[2mm]
\dfrac{v_i+h_i+i\omega_i}{\sqrt{2}}
\end{pmatrix},
\qquad i=1,2,
\end{align}
we obtain the quadratic terms in the charged, CP-odd, and CP-even sectors.
The quadratic part of the effective potential $V_{\rm eff}(\Phi_1,\Phi_2)$ can be written as
	\begin{align}
		V^{(2)}_{\rm eff}(\Phi_1,\Phi_2)\supset{}
		&\left(m_{12}^2+\frac{g_4^2v_1v_2}{4}\right)
			\begin{pmatrix}
			\phi_1^- & \phi_2^-
			\end{pmatrix}
				\begin{pmatrix}
				\frac{v_2}{v_1} & -1\\
				-1 & \frac{v_1}{v_2}
				\end{pmatrix}
			\begin{pmatrix}
			\phi_1^+\\
			\phi_2^+
			\end{pmatrix} \notag \\[2mm]
		&+\frac{1}{2} \left(m_{12}^2+g_4^2v_1v_2\right)
			\begin{pmatrix}
			\omega_1 & \omega_2
			\end{pmatrix}
				\begin{pmatrix}
				\frac{v_2}{v_1} & -1\\
				-1 & \frac{v_1}{v_2}
				\end{pmatrix}
			\begin{pmatrix}
			\omega_1\\
			\omega_2
			\end{pmatrix} \notag \\[2mm]
		&\quad +\frac{1}{2}
			\begin{pmatrix}
			h_1 & h_2
			\end{pmatrix}
				\begin{pmatrix}
				m_{11}^2 & -m_{12}^2\\
				-m_{12}^2 & m_{22}^2
				\end{pmatrix}
			\begin{pmatrix}
			h_1\\
			h_2
			\end{pmatrix}.
	\label{eq:higgs_quadratic_potential}
	\end{align}
The CP-even Higgs mass matrix is determined by the Hessian of the one-loop effective potential:
\begin{align}
\mathcal M_{\rm even}^2
&=
g_4^2R^2
\begin{pmatrix}
H_{11} & H_{12}\\
H_{12} & H_{22}
\end{pmatrix}
=
\begin{pmatrix}
m_{11}^2 & -m_{12}^2\\
-m_{12}^2 & m_{22}^2
\end{pmatrix},
\label{eq:CP_even_hessian}
\end{align}
where
\begin{align}
H_{ij}
=
\left.
\frac{\partial^2V_{\rm 1-loop}}
{\partial\alpha_i\partial\alpha_j}
\right|_{\alpha_1^{\rm min},\alpha_2^{\rm min}}.
\end{align}
The charged and CP-odd mass matrices are diagonalized by the angle $\beta$, defined as
\begin{align}
\tan\beta=\frac{v_2}{v_1}.
\end{align}
The corresponding mass eigenstates are given by
\begin{align}
\begin{aligned}
\begin{pmatrix}
G^+\\ H^+
\end{pmatrix}
=
\begin{pmatrix}
\cos\beta & \sin\beta\\
-\sin\beta & \cos\beta
\end{pmatrix}
\begin{pmatrix}
\phi_1^+\\ \phi_2^+
\end{pmatrix},
\\[3mm]
\begin{pmatrix}
G^0\\ A
\end{pmatrix}
=
\begin{pmatrix}
\cos\beta & \sin\beta\\
-\sin\beta & \cos\beta
\end{pmatrix}
\begin{pmatrix}
\omega_1\\ \omega_2
\end{pmatrix}.
\end{aligned}
\end{align}
The CP-even mass matrix is diagonalized by the mixing angle $\alpha$,
\begin{align}
\begin{pmatrix}
H\\ h
\end{pmatrix}
=
\begin{pmatrix}
\cos\alpha & \sin\alpha\\
-\sin\alpha & \cos\alpha
\end{pmatrix}
\begin{pmatrix}
h_1\\ h_2
\end{pmatrix},
\end{align}
where
\begin{align}
\tan 2\alpha
=
\frac{2m_{12}^2}{m_{11}^2-m_{22}^2}.
\label{eq:CP_even_mixing_angle}
\end{align}
The branch of $\alpha$ is chosen such that $m_h\leq m_H$.
The resulting mass eigenvalues are summarized as
\begin{align}
\text{Charged sector : }&\quad m_{G^\pm}^2 = 0, \qquad 
m_{H^\pm}^2 = \frac{g_4^2v^2}{4}+\frac{m_{12}^2}{\sin\beta\cos\beta},  \\[2mm]
\text{CP-odd sector : }&\quad m_{G^0}^2 = 0, \qquad 
m_A^2 = g_4^2v^2 +\frac{m_{12}^2}{\sin\beta\cos\beta},  \\[2mm]
\text{CP-even sector : }&
\begin{cases}
m_h^2 = \frac12\left[ m_{11}^2+m_{22}^2 -\sqrt{(m_{11}^2-m_{22}^2)^2+4m_{12}^4}\right], \\[4mm]
m_H^2= \frac12\left[ m_{11}^2+m_{22}^2 +\sqrt{(m_{11}^2-m_{22}^2)^2+4m_{12}^4} \right].
\end{cases}
\end{align}
Here, 
$h$ and $H$ denote the lighter and heavier CP-even Higgs bosons, respectively, while 
$G^\pm$ and $G^0$ are the Nambu--Goldstone bosons.
An interesting consequence of the present model is the mass relation between the charged and CP-odd Higgs bosons,
\begin{align}
m_A^2-m_{H^\pm}^2
= \frac34 g_4^2v^2
\label{eq:higgs_mass_relation}
\end{align}
This relation is independent of the radiatively generated quadratic mass parameters and follows from the tree-level quartic interactions fixed by the higher-dimensional gauge symmetry.

\begin{figure}[http]
\centering
\includegraphics[width=11.3cm]{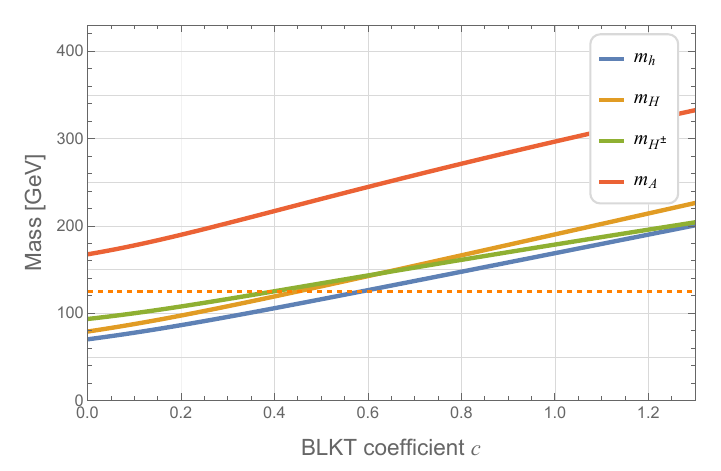}
\caption{
Physical Higgs boson masses as functions of the
BLKT coefficient $c$.
The blue, yellow, green, and red lines
represent $m_h$, $m_H$, $m_{H^\pm}$, and $m_A$,
respectively.
The horizontal dashed line indicates the
observed Higgs boson mass of 125 GeV.
}
\label{fig:higgs_mass_blkt}
\end{figure}

Finally, we examine the dependence of the physical Higgs boson masses on the BLKT coefficient $c$, as shown in Fig.~\ref{fig:higgs_mass_blkt}.
Although the BLKT has only a small effect on the shape of the one-loop effective potential, it modifies the relation between the compactification scale and the electroweak scale, thereby affecting the physical Higgs masses.
As $c$ increases, all four Higgs masses increase over the parameter range considered.
For $c=0.59$, the predicted mass of the SM-like Higgs boson is approximately $125$ GeV, consistent with the observed value.
At this parameter point, the Higgs mass spectrum is
given by
\begin{align}
\begin{aligned}
m_h &= 125.0\,\mathrm{GeV}, \qquad 
m_H = 140.9 \,\mathrm{GeV}, \\
m_{H^\pm} &= 142.0 \,\mathrm{GeV}, \qquad
m_A = 242.8\,\mathrm{GeV}.
\label{eq:benchmark_higgs_spectrum}
\end{aligned}
\end{align}

Thus, the observed Higgs boson mass can be accommodated by an appropriate choice of the BLKT coefficient, while the masses of the additional Higgs bosons are simultaneously determined within the model.

\subsection{Enhancement of the compactification scale by additional bulk scalars}

In the previous subsections, we have shown that the observed Higgs mass can be reproduced by adjusting the BLKT coefficient.
However, the compactification scale remains relatively low, $1/R \simeq 280$ GeV.
Such a low compactification scale is phenomenologically problematic because the KK excitations are expected to appear at low energies.
This motivates us to investigate whether the vacuum structure of the one-loop effective potential can be modified so as to obtain a higher compactification scale while maintaining the observed Higgs mass.
In particular, both small Wilson-line phase and sufficiently small curvature of the effective potential are important for this purpose.
As shown in \eqref{eq:fermion84_potential}, the contribution of the bulk fermion in the ${\bf{84}^*}$ representation is enhanced by its degrees of freedom.
We therefore introduce additional six-dimensional complex scalar fields in representations ${\cal R}$ of $SU(4)$, with the following bulk kinetic term:
\begin{equation}
 {\cal L}_{\rm scalar}
 =
 (D_M\Phi_{\cal R})^{\dagger}D^M\Phi_{\cal R}.
 \label{eq:additional_scalar_lagrangian}
\end{equation}
Their Wilson-line-dependent KK spectra contribute to the one-loop effective potential through Eq.~\eqref{eq:CW_potential}, with the bosonic statistics sign and two real degrees of freedom, $n_s=2$.
We consider scalar multiplets in the $\cal{R}={\bf 4^*},\ {\bf 10^*},\ {\bf 20^*},\ {\bf 35^*},\ {\bf 84^*}$, and adjoint representations.
The total one-loop effective potential is written as
\begin{equation}
 V_{1\text{-loop}}(\alpha_1,\alpha_2)
 =V_{1\text{-loop}}^{(g)} +V_{1\text{-loop}}^{(f_{84^*})} +\sum_{\cal R}N_{\cal R}  V_{1\text{-loop}}^{(s_{\cal R})}(\alpha_1,\alpha_2),
 \label{eq:full_potential_additional_scalars}
\end{equation}
where $N_{\cal R}$ denotes the number of complex-scalar multiplets.
Here the contribution of a single scalar multiplet is evaluated using its orbifold parities and $SU(4)$ representation.
We numerically determine the global minimum of Eq.~\eqref{eq:full_potential_additional_scalars} and the
curvature at that minimum. 
The BLKT coefficient $c$ is chosen to reproduce the SM Higgs mass, and the compactification scale is extracted from the lowest eigenvalue of the KK mass equation \eqref{eq:single_BLKT_KK}:
\begin{equation}
 \frac{1}{R}=\frac{m_W}{\xi_0}.
 \label{eq:compactification_scalar_scan}
\end{equation}
Table~\ref{tab:bulk_scalar_benchmarks} shows representative configurations.
We find that the compactification scale can be raised to the TeV range without introducing independent mass or quartic parameters into the Higgs potential.
In the next section, we investigate how the torus geometry affects the selected scalar configurations and their Higgs-sector predictions.

\begin{table}[http]
\centering
\caption{
Representative benchmark points for the enhancement of the compactification scale by additional bulk complex scalars. 
The integers denote the numbers of multiplets $N_{\mathcal R}$.
All benchmark points satisfy \(\alpha_1^{\min}=\alpha_2^{\min}=\alpha^{\rm min}\).
The BLKT coefficient $c$ is determined by requiring $m_h=125$ GeV, while $m_H$ is the heavier CP-even Higgs mass.}
 \label{tab:bulk_scalar_benchmarks}
\setlength{\tabcolsep}{4.8pt}
 \begin{tabular}{cccccc|ccccc}
 \hline
 & & & & &&&&&\\[-11pt]
${N_{{ 84}^{*}}}$
&$N_{{ 4}^{*}}$
&$N_{{ 10}^{*}}$
&$N_{{ 20}^{*}}$
&$N_{{ 35}^{*}}$
&$N_{{\text{ad}}}$
& $\alpha^{\rm min}$ 
& $c$ 
& $\xi_0$ 
& $1/R$ [TeV]
& $m_H$ [GeV]
\\
 &  & &&&&&&& \\[-11pt]
 \hline
11&0&0&0&2&0&0.11046&2.40&0.0419&1.91 & 498.3\\
11&0&2&1&2&0&0.06652&2.34&0.0256&3.13 & 722.0\\
11&0&0&2&2&0&0.05917&2.69&0.0217&3.69& 836.3\\
11&0&1&2&2&0&0.04989&2.81&0.0180&4.45& 943.4\\
11&2&0&0&3&0&0.04807&3.70&0.0156&5.12& 1121.4\\
11&0&0&0&3&2&0.02844&4.32&0.0087&9.21& 1558.8\\
\hline
\end{tabular}
\end{table}

\section{Torus Geometry and Higgs Phenomenology}
\label{sec:general_torus}

In the previous section, we investigated electroweak symmetry breaking and the Higgs mass spectrum on a square torus.
We now extend the analysis to a general torus characterized by the ratio of its radii and the angle between the two compact directions.
These additional geometric parameters modify the one-loop effective potential and the resulting Higgs mass spectrum.
We further examine the compatibility of the predicted scalar spectrum with electroweak precision measurements by evaluating its contribution to the oblique parameter $T$.

\subsection{Effects of the torus geometry on the effective potential}
\label{subsec:general_torus}
We generalize the compactification geometry from a square torus to a torus characterized by two radii $R_1$, $R_2$ and the angle $\theta$ between the compact directions.
Following Ref.~\cite{ABQ}, we introduce the complex structure modulus
\begin{align}
U=\frac{R_2}{R_1}e^{i\theta},
\qquad
u\equiv\frac{1}{\abs{U}}.
\end{align}
The square torus considered in the preceding sections corresponds to $u=1$ and $\theta=\pi/2$.
The metric of the compact space is given by
\begin{align}
G_{MN} &=
\begin{pmatrix}
R_1^2 & R_1R_2\cos\theta\\
R_1R_2\cos\theta & R_2^2
\end{pmatrix} \notag \\[3mm]
&=
R^2
\begin{pmatrix}
u^2 & u \cos\theta \\
u \cos\theta  & 1
\end{pmatrix}
\label{eq:general_torus_metric}
\end{align}
where $R_2 \equiv R$.
For a field with Wilson-line phases $\alpha_1$ and $\alpha_2$, the KK mass spectrum is generalized to
\begin{align}
M_{n_1,n_2}^2
&= \abs{\frac{(n_2+\alpha_2)-(n_1+\alpha_1)U}{\sqrt{\text{Im}\ U}{(\det G_{MN})}^{1/4}}}^2 \notag \\[2mm]
&=\frac{1}{R^2\sin^2\theta} \Big[ \frac{(n_1+\alpha_1)^2}{u^2} +(n_2+\alpha_2)^2 -2\frac{(n_1+\alpha_1)(n_2+\alpha_2)}{u} \cos\theta \Big].
\label{eq:general_torus_kk}
\end{align}
After Poisson resummation, the geometrical dependence of the Wilson-line effective potential is encoded in the winding-mode contributions
\begin{align}
\nu^\pm_{k_1,k_2} (\alpha_1,\alpha_2;u,\theta)
=\frac{ u\, \cos[q\pi(k_1\alpha_1\pm k_2\alpha_2)]\sin\theta}{ \left[(uk_1)^2+k_2^2 \pm2uk_1k_2\cos\theta \right]^3}.
\label{eq:general_torus_winding}
\end{align}
The one-loop effective potential retains the same combination of $SU(4)$ gauge and ${\bf{84^*}}$ fermion contributions as in the square-torus case, with the corresponding lattice sums generalized according to Eq.~\eqref{eq:general_torus_winding}.

The nonorthogonal compact directions also modify the normalization of the Higgs fields.
In terms of the Wilson-line phases $(\alpha_1, \alpha_2)$ and the canonically normalized vacuum expectation values
$(v_1, v_2)$, we write
\begin{align}
\begin{pmatrix}
\alpha_1 \\
\alpha_2
\end{pmatrix}
=g_4 R
L
\begin{pmatrix}
v_1 \\
v_2
\end{pmatrix}, 
\quad  
L\equiv
 \begin{pmatrix}
u & 0 \\
\cos \theta & \sin \theta
\end{pmatrix}.
\label{eq:general_wilson_vev}
\end{align}
Consequently, the electroweak vacuum expectation value satisfies
\begin{align}
v^2&= \frac{1}{g_4^2R^2\sin^2\theta}\qty(\frac{\alpha_1^2}{u^2}+\alpha_2^2 -2\frac{\alpha_1\alpha_2}{u}\cos\theta), \\[2mm]
\tan \beta &= \frac{u \alpha_2-\alpha_1 \cos \theta}{\alpha_1 \sin \theta}.
\label{eq:general_vev}
\end{align}
Thus, the geometrical parameters $u$ and $\theta$ affect the Higgs sector through both the one-loop effective potential and the normalization of the scalar fields.
The CP-even Higgs mass matrix is obtained from the curvature of the effective potential with respect to the Wilson-line phases.
Using the canonical normalization in Eq.~\eqref{eq:general_wilson_vev}, we obtain
\begin{align}
\mathcal M_{\rm even}^2
&=g_4^2R^2 L^{t} H L^{}  \notag \\[2mm]
&=g^2_4R^2
\begin{pmatrix}
u^2H_{11}+2u H_{12}\cos \theta+H_{22}\cos^2 \theta
& \left(uH_{12}+H_{22} \cos \theta \right)\sin \theta \\[2mm]
\left(uH_{12}+H_{22} \cos \theta \right)\sin \theta
&H_{22}\sin^2 \theta
\end{pmatrix}
\label{eq:general_even_mass}
\end{align}
The physical CP-even masses $m_h$ and $m_H$ are obtained by diagonalizing this matrix.
The CP-odd and charged Higgs masses are determined from the corresponding mass matrices, including the tree-level contributions originating from the higher-dimensional gauge interactions.

The couplings of the CP-even Higgs bosons to the electroweak gauge bosons are characterized by the mixing angles $\alpha$ and $\beta$.
Normalized to the corresponding Standard Model Higgs coupling, they are given by
\begin{align}
\kappa_{V}^h = \sin(\beta-\alpha),\qquad
\kappa_{V}^H = \cos(\beta-\alpha),
\qquad (V=W,Z).
\label{eq:higgs_gauge_couplings}
\end{align}
The alignment limit, in which the lighter CP-even Higgs boson has Standard Model-like gauge couplings, corresponds to $\beta-\alpha \simeq \pi/2$.

We now consider the equal-radius case $u=1$.
Owing to our orbifold parity assignments, the one-loop effective potential is invariant under the interchange of the two Wilson-line phases $(\alpha_1 \leftrightarrow \alpha_2)$.
Suppose that the vacuum is located along the symmetric direction $\alpha^{\rm min}_1=\alpha^{\rm min}_2 \ne 0$.
The exchange symmetry then requires the curvature to take the form
\begin{align}
H_{11}=H_{22}, \quad H_{12}=H_{21}.
\end{align}
The eigenvalues of the CP-even mass matrix are
\begin{align}
\begin{aligned}
m^2_+ &=g_4^2 R^2 (H_{11}+H_{12})(1+\cos\theta), \\
m^2_- &=g_4^2 R^2(H_{11}-H_{12})(1-\cos\theta).
\end{aligned}
\end{align}
The corresponding normalized eigenvectors are
\begin{align}
\hat{\boldsymbol{e}}_{+} = 
\begin{pmatrix}
\cos \dfrac{\theta}{2} \\[3mm]
\sin \dfrac{\theta}{2} 
\end{pmatrix},\quad 
\hat{\boldsymbol{e}}_{-} =
\begin{pmatrix}
-\sin \dfrac{\theta}{2} \\[3mm]
\cos \dfrac{\theta}{2}
\end{pmatrix}.
\end{align}
On the other hand, the normalized vacuum direction follows from Eq.~\eqref{eq:general_wilson_vev}.
For $u=1$ and $\alpha_1^{\rm min}=\alpha_2^{\rm min} \ne 0 $, we obtain
\begin{align}
\hat{\boldsymbol v} = 
\begin{pmatrix}
\cos \dfrac{\theta}{2} \\[3mm]
\sin \dfrac{\theta}{2} 
\end{pmatrix},
\end{align}
which also implies
\begin{align}
\tan\beta=\tan\frac{\theta}{2}.
\end{align}
The couplings of the CP-even Higgs bosons to the electroweak gauge bosons, normalized to the corresponding SM couplings, are determined by the inner products between their mass eigenvectors and the vacuum direction,
\begin{align}
\kappa^{+} =
\left| \hat{\boldsymbol e}_+ \cdot \hat{\boldsymbol v} \right|=1,
\qquad
\kappa^{-} =
\left| \hat{\boldsymbol e}_- \cdot \hat{\boldsymbol v} \right|=0.
\end{align}
Thus, the vacuum direction is an eigenvector of the CP-even Higgs mass matrix for any $0< \theta < \pi$.
The identification of the SM-like Higgs boson depends on the ordering of the two CP-even mass eigenvalues. 
For $m_+^2<m_-^2$, the lighter Higgs boson is aligned with the vacuum direction,
\begin{align}
\kappa_V^h=1,\qquad \kappa_V^H=0.
\end{align}
This case is realized when
\begin{align}
H_{12}+H_{11}\cos\theta<0.
\end{align}
Conversely,  for $m_-^2<m_+^2$, the heavier CP-even Higgs boson mass is aligned with the vacuum direction.

This establishes exact alignment as a consequence of the exchange symmetry and the symmetric vacuum configuration, without requiring a decoupling limit for the additional Higgs bosons.
We emphasize that the exchange symmetry alone does not guarantee that the global minimum lies along the symmetric direction. 
Moreover, the ordering of the two CP-even mass eigenvalues depends on the curvature of the effective potential and the torus angle.
These conditions must therefore be examined in the numerical analysis.

\subsection{Mass spectrum and experimental constraints}
\label{subsec:general_higgs_spectrum}
We now examine how the torus geometry affects the Higgs mass spectrum and the couplings of the CP-even Higgs bosons to electroweak gauge bosons.
Mass splittings among the additional Higgs bosons can induce custodial-symmetry-breaking contributions to electroweak precision
observables~\cite{BBW,GLOO}. 
Deviations from the alignment limit can also be probed through charged-Higgs processes such as $H^\pm\to W^\pm h$~\cite{Hashemi}.

\begin{figure}[H]
  \centering
  \includegraphics[width=0.6\linewidth]
    {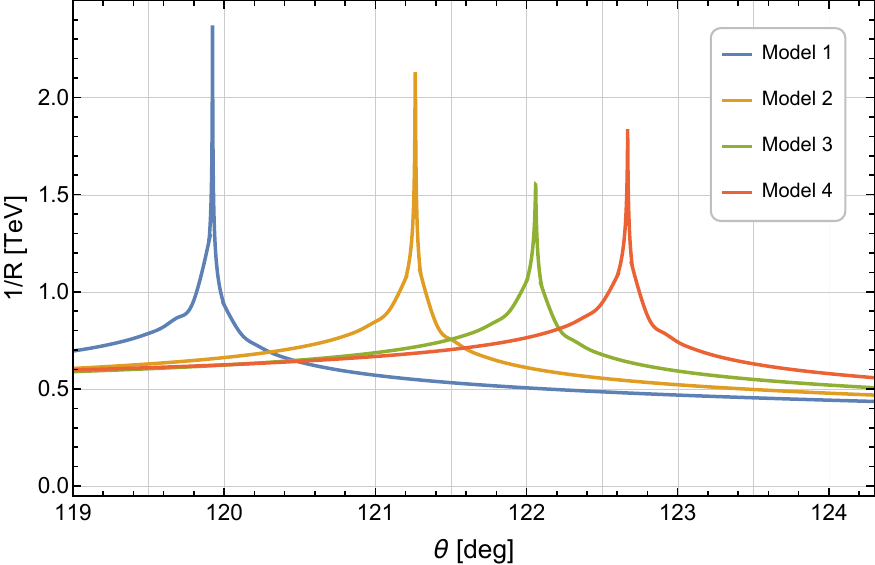}
  \caption{
Dependence of the compactification scale $1/R$ on the torus angle $\theta$ for $u=1$.
The BLKT coefficient $c$ is adjusted at each angle to reproduce $m_h\simeq125.1~\mathrm{GeV}$.}
  \label{fig:compactification_theta}
\end{figure}

We first focus on the equal-radius case, $u=1$.
For the four benchmark models studied below, the minimum of the one-loop effective potential lies along the symmetric direction,
$\alpha_1^{\rm min}=\alpha_2^{\rm min}\neq 0$, over the range of torus angles considered. At these minima, the lighter CP-even Higgs boson exhibits exact alignment, $\kappa_V^h=1$, in agreement with the analytical result derived in the preceding subsection.
The benchmark models are specified by the bulk scalar multiplicities
$ {\bf{N}}_{\mathcal{R}}=(N_{84}, N_{4},N_{10},N_{20},N_{35},N_{\text{ad}})$, which are summarized in Table~\ref{tab:benchmark_models}.
For each model and torus angle, the BLKT coefficient $c$ is adjusted to reproduce $m_h \simeq 125.1~\mathrm{GeV}$.
Figure~\ref{fig:compactification_theta} shows the dependence of the compactification scale $1/R$ on the torus angle $\theta$.
Near these peaks, the compactification scales range from $1.55$ to $2.37~\mathrm{TeV}$, as summarized in Table~\ref{tab:benchmark_models}.
Thus, compactification scales of order $2~\mathrm{TeV}$ can be realized in different scalar configurations while reproducing
the observed Higgs mass and maintaining exact alignment.
The additional Higgs masses lie approximately in the range $1.7$--$4.4~\mathrm{TeV}$.
Despite these large masses, the splitting between $H$ and $H^\pm$ remains relatively small, $|m_H-m_{H^\pm}|=26$--$32~\mathrm{GeV}$, for the four benchmark points.
Such a near-degeneracy tends to suppress custodial-symmetry-breaking contributions from the additional scalar sector.

\begin{table}[htbp]
\centering
\caption{Bulk scalar multiplicities and the corresponding compactification scales and additional Higgs masses near the peaks in Fig.~\ref{fig:compactification_theta} for $u=1$.
The angle $\theta_0$ denotes the peak position for each model.}
 \label{tab:benchmark_models}
\setlength{\tabcolsep}{3.1pt}
 \begin{tabular}{c|cccccc}
& scalar multiplet & $\theta_0$ & $1/R$ [TeV]& $m_H$ [GeV]& $m_{H^\pm}$  [GeV] & $m_{A}$  [GeV] \\ 
 \hline
&&&&&&\\[-11pt]
Model 1 & ${\bf{N}}_{\cal{R}}=(4,16,2,0,2,0)$ & $ 119.923^\circ$ & $ 2.37$ &4261 &4287 &4369 \\[1mm]
Model 2 & ${\bf{N}}_{\cal{R}}=(5,13,1,2,0,0)$ & $ 121.264^\circ$ & $ 2.13$ &3403 &3430 &3517 \\[1mm]
Model 3 & ${\bf{N}}_{\cal{R}}=(6,11,0,1,0,1)$ & $ 122.060^\circ$ & $ 1.55$ &1712 &1740 &1834 \\[1mm]
Model 4 & ${\bf{N}}_{\cal{R}}=(7,9,2,0,0,0)$ & $ 122.669^\circ$ & $ 1.84$ &2240 &2272 &2376 \\[1mm]
\end{tabular}
\end{table}

\begin{figure}[t]
  \centering
  \includegraphics[width=0.6\linewidth]{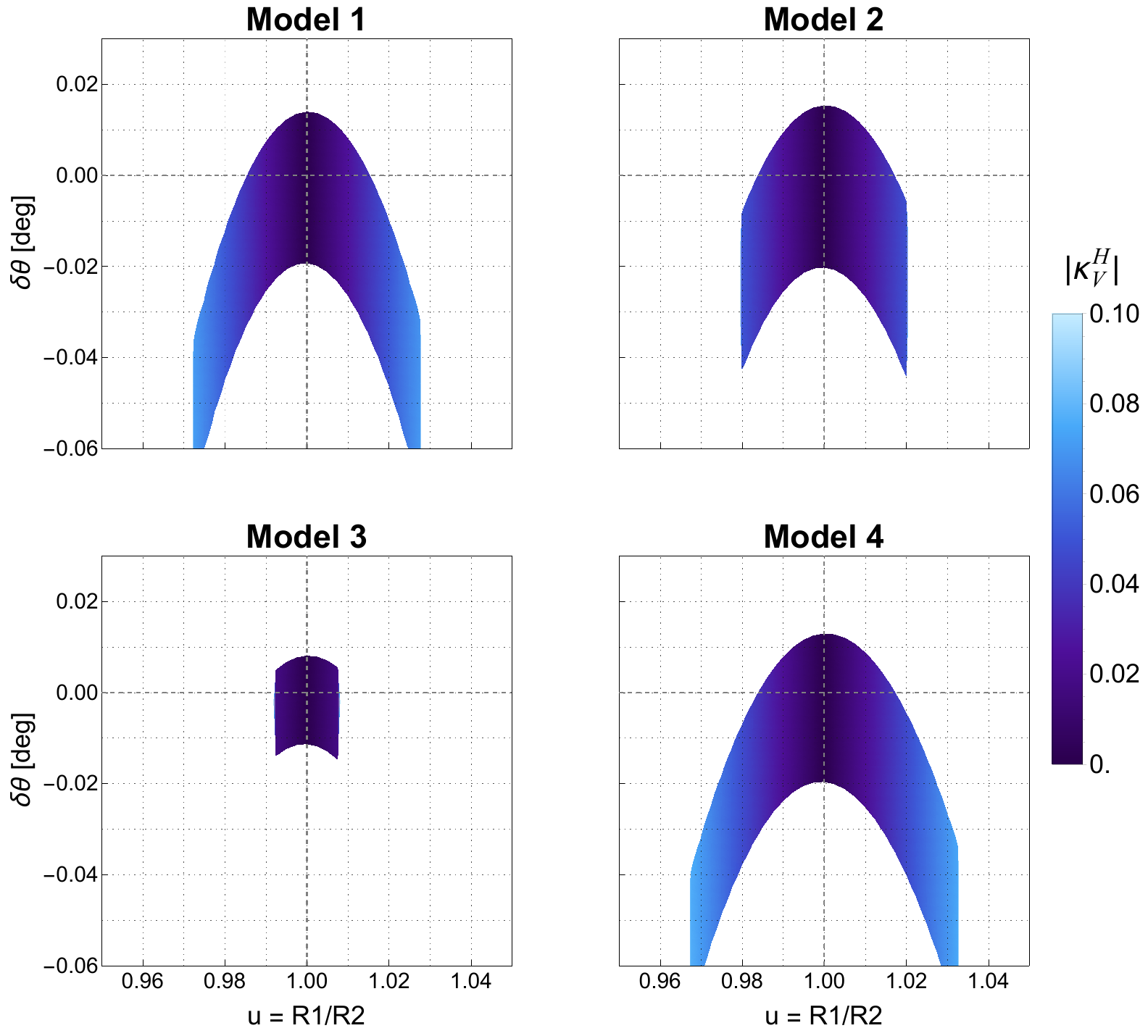}
  \caption{Regions satisfying the alignment condition and compactification-scale criteria in the $(u,\delta\theta)$ plane for the four benchmark models, where $\delta\theta=\theta-\theta_0$, with $\theta_0$ denotes the peak angle listed in Table~\ref{tab:benchmark_models}.
The color scale represents the magnitude of the normalized heavy CP-even Higgs coupling to electroweak gauge bosons, $\kappa_V^H$, evaluated at the global minimum of the one-loop effective potential.
Only points satisfying $|\kappa_V^H|<0.1$ and $1/R > 1.3~\mathrm{TeV}$ are shown.}
  \label{fig:alignment_plane}
\end{figure}

We next extend the analysis to general torus geometries by varying both $u$ and $\theta$ for the four benchmark models.
For each point in the $(u,\theta)$ plane, we determine the minimum of the one-loop effective potential and evaluate the physical Higgs spectrum and the Higgs--gauge-boson couplings.
As discussed in the preceding subsection, exact alignment is realized at a symmetric vacuum for $u=1$ when the aligned state is the lighter CP-even Higgs boson.
For $u\neq 1$, the exchange symmetry between the Wilson-line phases is generally lost, so exact alignment is no longer guaranteed.
We therefore focus on the parameter region satisfying $|\kappa_V^H| < {0.100}$, which corresponds to $\kappa_V^h \gtrsim 0.995$.
We additionally impose $1/R > 1.3\ \mathrm{TeV}$ as a numerical selection criterion to focus on compactification scales in the TeV range.
Figure~\ref{fig:alignment_plane} shows the corresponding regions in the $(u,\delta\theta)$ plane, where $\delta\theta \equiv \theta - \theta_0$ with $\theta_0$ denotes in Table~\ref{tab:benchmark_models}.
All four models admit finite regions away from the equal-radius limit that satisfy both criteria.
The extent of these regions depends on the scalar content: Models 1 and 4 allow relatively broad deviations from $u=1$, whereas Model 3 is restricted to a much narrower range.
Thus, approximate alignment can be maintained for unequal torus radii, although the allowed geometric deformations are model dependent.

\begin{figure}[H]
  \centering
  \includegraphics[width=0.90\linewidth]{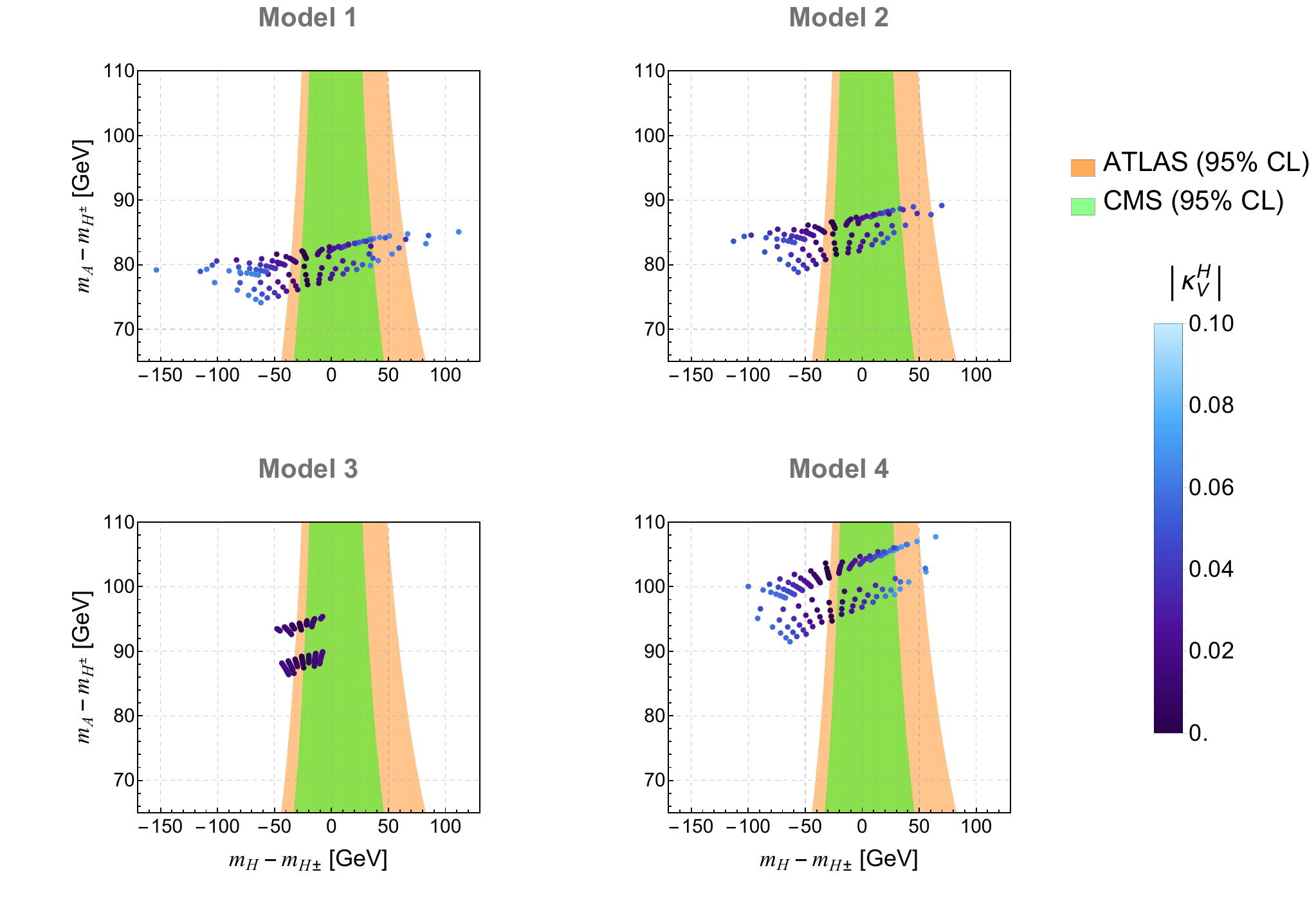}
\caption{Predicted mass splittings $\Delta m_H = m_H-m_{H^\pm}$ and $\Delta m_A = m_A-m_{H^\pm}$ for the four benchmark models.
Each point corresponds to a parameter set in the $(u,\theta)$ plane satisfying $|\kappa_V^H| < 0.1$ and $1/R > 1.3~\mathrm{TeV}$.
The points are colored according to $|\kappa_V^H|$.
The shaded bands represent the 95\% confidence regions inferred from the $W$-boson mass measurements \cite{ATLASW2024,CMS2026} within the $\Delta\rho$ approximation.
For each parameter point, the global minimum of the effective potential is used, and the BLKT coefficient $c$ is adjusted to reproduce $m_h=125.1~\mathrm{GeV}$.}
\label{fig:W_constraints}
\end{figure}

We further examine the additional Higgs mass splittings in relation to electroweak precision measurements.
In the 2HDM, mass splittings among the additional Higgs bosons induce radiative corrections to the electroweak $\rho$ parameter and
thereby affect the prediction for the $W$-boson mass~\cite{BBW,GLOO}.
We define
\begin{align}
\Delta m_H = m_H-m_{H^\pm}, \qquad 
\Delta m_A = m_A-m_{H^\pm}.
\label{eq:scalar_mass_splittings}
\end{align}
Figure~\ref{fig:W_constraints} shows the predicted mass splittings in the $(\Delta m_H,\Delta m_A)$ plane.
The shaded bands denote the 95\% confidence regions inferred from the ATLAS and CMS measurements of the $W$-boson mass within the scalar-sector $\Delta\rho$ approximation \cite{ATLASW2024,CMS2026}.
The comparison is performed within the $\Delta\rho$ approximation, focusing on the contributions from the additional scalar bosons.
Contributions from KK excitations and other possible new-physics effects are not included in this comparison.
For all four models, $\Delta m_A$ is positive and typically of order $100~\mathrm{GeV}$, while $\Delta m_H$ spans a wider range and can approach zero.
The predicted points partially overlap with the regions inferred from the ATLAS and CMS measurements.
The degree of overlap depends on the benchmark model, reflecting the dependence of the scalar spectrum on the bulk field content and torus geometry.

The above results demonstrate that the torus geometry provides an important handle on the compactification scale, Higgs alignment, and the additional scalar spectrum in six-dimensional GHU.
In the equal-radius case, all four benchmark models realize compactification scales of order $2~\mathrm{TeV}$ together with exact alignment at the potential minimum.
Away from this limit, finite parameter regions retain approximately SM-like couplings of the lighter Higgs boson and compactification scales above $1.3~\mathrm{TeV}$.
Within the scalar-sector $\Delta\rho$ approximation, these regions also contain spectra compatible with the ATLAS and CMS $W$-boson mass measurements.
A complete electroweak precision analysis, including KK contributions, is left for future investigation.

\section{Conclusions}

In this paper, we have investigated electroweak symmetry breaking and the Higgs mass spectrum in a six-dimensional $SU(4)$ gauge--Higgs unification model compactified on $T^2/Z_2$, including brane-localized gauge kinetic terms (BLKTs).
The two Higgs doublets arise from the extra-dimensional components of the gauge field, and their interactions are constrained by the higher-dimensional gauge symmetry.
In particular, the tree-level Higgs potential is CP conserving and $Z_2$ symmetric, while the gauge structure predicts $\sin^2\theta_W=1/4$ at the compactification scale.
The fermion representations also reflect the characteristic features of the $SU(4)$ gauge structure.
The fundamental representation accommodates electric charge assignments compatible with those of quarks, while its conjugate representation yields chiral zero modes corresponding to the left-handed quark doublet and the right-handed up-type quark.
Furthermore, the matrix elements of the gauge generators in the covariant derivative for the sixth-rank totally symmetric representation $\mathbf{84}^*$ provide an enhanced Yukawa coupling, yielding the tree-level mass relation $m_t\simeq 2m_W$ for the top quark.
The Yukawa interactions of bulk fermions are naturally aligned, thereby suppressing tree-level flavor-changing neutral currents in the up-type quark sector.

We have derived the KK mass eigenvalue equations in the presence of BLKTs and examined their effects on the gauge boson spectrum.
By evaluating the one-loop effective potential, including contributions from the $SU(4)$ gauge field, the bulk fermion in the $\mathbf{84^*}$ representation, and additional bulk scalar multiplets, we have shown that electroweak symmetry breaking can be realized while reproducing the observed Higgs boson mass of approximately $125~\mathrm{GeV}$.
The introduction of additional bulk scalars allows the compactification scale to be raised to the TeV range, while the masses of the additional Higgs bosons are determined by the resulting effective potential and the tree-level gauge interactions.

We have further extended the analysis to a general torus characterized by the radius ratio $u=R_1/R_2$ and the angle $\theta$.
These geometrical parameters modify the KK spectrum, the one-loop effective potential, and the relation between the Wilson-line phases and the canonically normalized vacuum expectation values of the Higgs fields.
In the equal-radius case, $u=1$, we have demonstrated analytically that exact alignment follows from the exchange symmetry of the Wilson-line phases when the vacuum lies along the symmetric direction and the aligned CP-even state is the lighter Higgs boson.
For four benchmark models, we have found symmetric global minima that realize exact alignment together with compactification scales ranging from $1.55$ to $2.37~\mathrm{TeV}$.
For general torus geometries with $u\neq1$, we have identified finite parameter regions satisfying $|\kappa_V^H |<0.1$ and $1/R>1.3~\mathrm{TeV}$.
We have also examined the additional Higgs mass splittings in light of precision measurements of the $W$-boson mass.
Within the scalar-sector $\Delta\rho$ approximation, we have identified parameter regions in which the predicted mass spectra are compatible with the ATLAS and CMS measurements.
These results indicate that compactification scales in the TeV range, approximately SM-like Higgs gauge couplings, and additional scalar spectra consistent with the electroweak precision constraints considered here can be realized simultaneously.

A complete phenomenological assessment requires further investigation of the contributions from KK excitations to electroweak precision observables, as well as constraints from direct searches for additional Higgs bosons.
Nevertheless, our results demonstrate that the bulk matter content, BLKTs, and compactification geometry provide complementary mechanisms for realizing a phenomenologically promising Higgs sector within six-dimensional gauge--Higgs unification.

\subsection*{Acknowledgments}
This work was supported by Japan Society for the Promotion of Science, Grants-in-Aid for Scientific Research, No. JP25K07304 (N.M.), 
and JST SPRING, Grant Number JPMJSP2139 (A.N.), 
and KSU Fundamental Research Fund in Kyushu Sangyo University, T024023 (T.H.).

\vspace*{10mm}

\appendix
\section{\mbox{\boldmath $SU(4)$} generators}
\label{app:generator}
The $SU(4)$ generators $T^a=t^a/2$ are listed below. 
	\begin{align*}
		t^1&=
			\begin{pmatrix}\mqty{
			0 & 1 & 0 &0 \\
			1 & 0 & 0 &0 \\
			0 & 0 & 0 &0 \\
			0 & 0 & 0 &0 
			}\end{pmatrix},
		\;
		t^2=
			\begin{pmatrix}\mqty{
			0 & -i & 0 &0 \\
			i & 0 & 0 &0 \\
			0 & 0 & 0 &0 \\
			0 & 0 & 0 &0 
			}\end{pmatrix},
		\;
		t^3=
			\begin{pmatrix}\mqty{
			1 & 0 & 0 &0 \\
			0 & -1 & 0 &0 \\
			0 & 0 & 0 &0 \\
			0 & 0 & 0 &0 
			}\end{pmatrix},\\
		t^4&=
			\begin{pmatrix}\mqty{
			0 & 0 & 1 &0 \\
			0 & 0 & 0 &0 \\
			1 & 0 & 0 &0 \\
			0 & 0 & 0 &0 
			}\end{pmatrix},
		\;
		t^5=
			\begin{pmatrix}\mqty{
			0 & 0 & -i &0 \\
			0 & 0 & 0 &0 \\
			i & 0 & 0 &0 \\
			0 & 0 & 0 &0 
			}\end{pmatrix},\\
		t^6&=
			\begin{pmatrix}\mqty{
			0 & 0 & 0 &0 \\
			0 & 0 & 1 &0 \\
			0 & 1 & 0 &0 \\
			0 & 0 & 0 &0 
			}\end{pmatrix},
		\;
		t^7=
			\begin{pmatrix}\mqty{
			0 & 0 & 0 &0 \\
			0 & 0 & -i &0 \\
			0 & i & 0 &0 \\
			0 & 0 & 0 &0 
			}\end{pmatrix},
		\;
		t^8=\frac{1}{\sqrt{3}}
			\begin{pmatrix}\mqty{
			1 & 0 & 0 &0 \\
			0 & 1 & 0 &0 \\
			0 & 0 & -2 &0 \\
			0 & 0 & 0 &0 
			}\end{pmatrix},\\
		t^9&=
			\begin{pmatrix}\mqty{
			0 & 0 & 0 &1 \\
			0 & 0 & 0 &0 \\
			0 & 0 & 0 &0 \\
			1 & 0 & 0 &0 
			}\end{pmatrix},
		t^{10}=
			\begin{pmatrix}\mqty{
			0 & 0 & 0 &-i \\
			0 & 0 & 0 &0 \\
			0 & 0 & 0 &0 \\
			i & 0 & 0 &0 
			}\end{pmatrix},
		t^{11}=
			\begin{pmatrix}\mqty{
			0 & 0 & 0 &0 \\
			0 & 0 & 0 &1 \\
			0 & 0 & 0 &0 \\
			0 & 1 & 0 &0 
			}\end{pmatrix},
		t^{12}=
			\begin{pmatrix}\mqty{
			0 & 0 & 0 &0 \\
			0 & 0 & 0 &-i \\
			0 & 0 & 0 &0 \\
			0 & i & 0 &0 
			}\end{pmatrix},\\
		t^{13}&=
			\begin{pmatrix}\mqty{
			0 & 0 & 0 &0 \\
			0 & 0 & 0 &0 \\
			0 & 0 & 0 &1 \\
			0 & 0 & 1 &0 
			}\end{pmatrix},
		t^{14}=
			\begin{pmatrix}\mqty{
			0 & 0 & 0 &0 \\
			0 & 0 & 0 &0 \\
			0 & 0 & 0 &-i \\
			0 & 0 & i &0 
			}\end{pmatrix},
		t^{15}=\frac{1}{\sqrt{6}}
			\begin{pmatrix}\mqty{
			1 & 0 & 0 &0 \\
			0 & 1 & 0 &0 \\
			0 & 0 & 1 &0 \\
			0 & 0 & 0 &-3 
			} \end{pmatrix}.
	\end{align*}
These generators satisfy the orthonormality relation, $\text{Tr}[T^aT^b]=\delta_{ab}/2$.
	
\section{Fermion representation and KK spectrum}
\label{app:fermion_representation}
We summarize the construction of the totally symmetric tensor representations of $SU(4)$ and their couplings to the Wilson-line background.  
First, we consider a fermion field belonging to an $N$-rank totally symmetric tensor representation of $SU(4)$.
For an $N$-rank totally symmetric tensor $\Psi_{i_1\cdots i_N}$ , the dimension of  the representation is
\begin{equation}
 \dim[\Psi_{i_1\cdots i_N}]=\frac{(N+3)(N+2)(N+1)}{6}.
\end{equation}
Under an $SU(4)$ gauge transformation, a field in the fundamental representation transforms as
\begin{align}
\psi_i \to U_i{}^j \psi_j,
\end{align}
where $i,j$ are $SU(4)$ indices.
Accordingly, the transformation for an $N$-rank totally symmetric tensor field $\Psi_{i_1\cdots i_N}$ is
\begin{align}
\Psi_{i_1\cdots i_N}\rightarrow
U_{i_1}{}^{j_1}\cdots U_{i_N}{}^{j_N} \Psi_{j_1\cdots j_N}.
\end{align}
Thus, the covariant derivative is given by
\begin{align}
D_M\Psi_{i_1\cdots i_N}
=
\partial_M\Psi_{i_1\cdots i_N}+ig_6 A_M^a \sum_{k=1}^{N}
(T_{\bf{4}}^a)_{i_k}{}^j \Psi_{i_1\cdots i_{k-1}ji_{k+1}\cdots i_N},
 \label{eq:appB-cov}
\end{align}
where $g_6$ is the six-dimensional gauge coupling and $T_{\bf{4}}^a$ are the generators of $SU(4)$ in the fundamental representation.
To evaluate the matrix elements of the generators, we decompose the fundamental representation under $SU(2)_L\times U(1)_Y$ as
\begin{equation}
 {\bf4}\longrightarrow D_{-1/6}\oplus A_{+1}\oplus B_{-2/3},
 \label{eq:appB-fund}
\end{equation}
where $D={\bf{2}},\ A=B={\bf{1}}$ denote $SU(2)_L$ representations and subscripts denote the eigenvalues of $Y/2$.
A component of the symmetric representation is denoted by $\Psi_m^{(d,a,b)}$, corresponding to $D^d \otimes A^a \otimes B^b$, where $d+a+b=N$ and $m=-d/2,\ldots,d/2$ is the third component of weak isospin.

Introducing occupation numbers $n_\pm=d/2\pm m$, the normalized states are
\begin{equation}
\Psi_m^{(d,a,b)}
=\frac{(a_+^\dagger)^{n_+}(a_-^\dagger)^{n_-}
(a_A^\dagger)^a(a_B^\dagger)^b}{\sqrt{n_+!n_-!a!b!}} \ \Psi_0^{(0,0,0)}.
\end{equation}
These creation operators act on the symmetric tensor indices and should not be confused with physical fields.
The neutral Higgs background connects the upper component of the $SU(2)_L$ doublet $D$ and the singlet $B$.
We introduce the operators
\begin{equation}
E_B=a_B^\dagger a_+.
\end{equation}
Its action on the normalized states is
\begin{align}
\begin{aligned}
E_B\Psi_m^{(d,a,b)}={}&\sqrt{\left(\frac d2+m\right)(b+1)}\Psi_{m-1/2}^{(d-1,a,b+1)},
\end{aligned}
\end{align}
The Higgs generator is therefore $T_H=(E_B+E_B^\dagger)/2$.
As an example, we consider the sixth-rank totally symmetric representation $\mathbf{84}$.
This representation is decomposed under $SU(2)_L\times U(1)_Y$ as
\begin{align}
\begin{aligned}
\mathbf{84}\rightarrow{}&
\mathbf{7}_{-1}
+\mathbf{6}_{\frac16}
+\mathbf{6}_{-\frac32}
+\mathbf{5}_{\frac43}
+\mathbf{5}_{-\frac13}
+\mathbf{5}_{-2}
\\
&+\mathbf{4}_{\frac52}
+\mathbf{4}_{\frac56}
+\mathbf{4}_{-\frac56}
+\mathbf{4}_{-\frac52}
+\mathbf{3}_{\frac{11}{3}}
+\mathbf{3}_{2}
+\mathbf{3}_{\frac13}
+\mathbf{3}_{-\frac43}
+\mathbf{3}_{-3}
\\
&\quad +\mathbf{2}_{\frac{29}{6}}
+\mathbf{2}_{\frac{19}{6}}
+\mathbf{2}_{\frac32}
+\mathbf{2}_{-\frac16}
+\mathbf{2}_{-\frac{11}{6}}
+\mathbf{2}_{-\frac72}
\\
&\qquad +\mathbf{1}_{6}
+\mathbf{1}_{\frac{13}{3}}
+\mathbf{1}_{\frac83}
+\mathbf{1}_{1}
+\mathbf{1}_{-\frac23}
+\mathbf{1}_{-\frac73}
+\mathbf{1}_{-4}.
\label{eq:appB_decomposition}
\end{aligned}
\end{align}
The relevant components are
\begin{equation}
Q_L\sim\Psi_{1/2}^{(1,2,3)}, \qquad u_R\sim \Psi_0^{(0,2,4)}.
\label{eq:appB_top}
\end{equation}
The corresponding matrix element is
\begin{equation}
E_B\Psi_{1/2}^{(1,2,3)}= 2\Psi_0^{(0,2,4)}.
\label{eq:appB_top_matrix}
\end{equation}

To obtain the KK mass spectrum, we consider the matrix elements of the Higgs generator $T_H$.
Since $T_H$ changes only $n_+$ and $b$, the occupation numbers $a$ and $n_-$ are conserved.
We therefore introduce
\begin{equation}
K=a+n_-.
\label{eq:appB_L}
\end{equation}
For a fixed pair $(a,n_-)$, the remaining occupation numbers satisfy
\begin{equation}
n_+ + b=6-K.
\label{eq:appB_constraint}
\end{equation}
Consequently, the states connected by $T_H$ form a block of dimension $\ell = 7-K, (K=0,1, \dots 6)$.
For each $K$, there are $K+1$ possible combinations of $(a,n_-)$.
The orbifold parity $(p_0,p_2)$ of a state is determined by those of the fundamental components,
\begin{align}
D:(+,+),\qquad
A:(+,-),\qquad
B:(-,-).
\end{align}
Thus, the complete representation is decomposed into blocks as
\begin{align}
\begin{aligned}
\mathbf{84}\rightarrow{}&
\mathbf7_{(++)} \oplus \mathbf6_{(++)} \oplus \mathbf6_{(+-)} \oplus 2 \times \mathbf5_{(++)} \oplus \mathbf5_{(+-)} \\
&\oplus 2 \times \mathbf4_{(++)} \oplus 2 \times \mathbf4_{(+-)} \oplus 3 \times \mathbf3_{(++)} \oplus 2 \times \mathbf3_{(+-)}
\\
&\oplus 3 \times \mathbf2_{(++)} \oplus 3  \times \mathbf2_{(+-)} \oplus4 \times \mathbf1_{(++)} \oplus3 \times \mathbf1_{(+-)}.
\end{aligned}
\end{align}
Here, the boldface numbers indicate the block dimensions, and the coefficients indicate their multiplicities.
For a block of dimension $\ell$, the nonvanishing
matrix elements of $T_H$ are
\begin{equation}
(T_H^{(\ell)})_{k,k+1}
=(T_H^{(\ell)})_{k+1,k}
=\frac12\sqrt{k(\ell-k)}, \qquad k=1,\ldots,\ell-1.
\end{equation}
This matrix is equivalent to the $SU(2)$ generator
$J_x$ with spin $j=(\ell-1)/2$.
Its eigenvalues are
\begin{equation}
\lambda=-j,-j+1,\ldots,j.
\end{equation}

The mass-squared operator is obtained from
$\widehat M^2=-(D_5^2+D_6^2)$.
Using the Wilson-line phases $\alpha_1,\alpha_2$,
we obtain 
\begin{align}
\widehat M^2_{n_1,n_2}
&=\frac{1}{R^2}
\Big[
\qty{\qty(n_1+\frac{\varepsilon}{2})^2+\qty(n_2+\frac{\varepsilon}{2})^2 }I_N+(\alpha^2_1+\alpha^2_2) \qty(T^{(\ell)}_H )^2 \notag \\
&\hspace{5.cm} +2\qty{\qty(n_1+\frac{\varepsilon}{2})\alpha_1+\qty(n_2+\frac{\varepsilon}{2})\alpha_2 }T^{(\ell)}_H \Big],
\end{align}
and mass eigenvalues
\begin{align}
&M^2_{n_1,n_2}=\frac{1}{R^2}
\left[\qty(n_1+\frac{\varepsilon}{2}+\lambda\alpha_1)^2+\qty(n_2+\frac{\varepsilon}{2}+\lambda\alpha_2)^2 \right],
\end{align}
where $n_1,n_2\in\mathbb Z$ and $\varepsilon=0$ or $1$ for $(++)$ or  $(+-)$ parity assignments, respectively.
The KK mass spectrum of each block of {\bf{84}} is determined by the eigenvalues $\lambda$ and boundary-condition parameter $\epsilon$ summarized in Table~\ref{tab:fermion_KK}.
\begin{table}[http]
\centering
\caption{Wilson-line eigenvalues and boundary conditions
for each block of the $\mathbf{84}$ representation.}
\label{tab:fermion_KK}
\begin{tabular}{ccc}
Block & $\lambda$ & $\varepsilon$ \\ 
\hline
&& \\[-11pt]
$\mathbf{7}_{(++)}$ & $0,\pm1,\pm2,\pm3$ & $0$ \\[2mm]
$\mathbf{6}_{(++)}$ & $\pm\frac12,\pm\frac32,\pm\frac52$ & $0$ \\[2mm]
$\mathbf{6}_{(+-)}$ & $\pm\frac12,\pm\frac32,\pm\frac52$ & $1$ \\[2mm]
$\mathbf{5}_{(++)}$ & $0,\pm1,\pm2$ & $0$ \\[2mm]
$\mathbf{5}_{(+-)}$ & $0,\pm1,\pm2$ & $1$ \\[2mm]
$\mathbf{4}_{(++)}$ & $\pm\frac12,\pm\frac32$ & $0$ \\[2mm]
$\mathbf{4}_{(+-)}$ & $\pm\frac12,\pm\frac32$ & $1$ \\[2mm]
$\mathbf{3}_{(++)}$ & $0,\pm1$ & $0$ \\[2mm]
$\mathbf{3}_{(+-)}$ & $0,\pm1$ & $1$ \\[2mm]
$\mathbf{2}_{(++)}$ & $\pm\frac12$ & $0$\\[2mm]
$\mathbf{2}_{(+-)}$ & $\pm\frac12$ & $1$
\end{tabular}
\end{table}

\newpage
\section{KK mass equation with BLKTs}
\label{app:BLKT_equation}
We derive the KK mass equation in the presence of brane-localized kinetic terms (BLKTs) on $T^2/Z_2$ and discuss the properties of its solutions.
The four fixed points are denoted by ${\bf{y}}_i=(x_i^5,x_i^6)$, where
\begin{equation}
\begin{aligned}
{\bf y}_0&=(0,0),&
{\bf y}_1&=(\pi R,0),\\
{\bf y}_2&=(0,\pi R),&
{\bf y}_3&=(\pi R,\pi R).
\end{aligned}
\end{equation}
We consider the six-dimensional Lagrangian of gauge field
\begin{align}
\mathcal L_{\rm{gauge}}
={}&-\frac14 F^a_{MN}F^{aMN}
\nonumber\\
&-\frac14
\sum_{i=0}^{3} 2\pi^2 R^2 c_i
\delta^{(2)}({\bf{y}}-{\bf{y}}_i)
F^b_{\mu\nu}F^{b\mu\nu},
\label{eq:appC_L}
\end{align}
where $c_i$ are dimensionless BLKT coefficients, and the index $b$ denotes the gauge components to which the localized operators couple.

After diagonalizing the Wilson-line background, the extra-dimensional wave function $f(\bf y)$ satisfies
\begin{equation}
\left[ \partial_5^2+\partial_6^2+ M^2\left\{ 1+2\pi^2 R^2 \sum_{i=0}^{3}c_i \delta^{(2)}({\bf{y}}-{\bf{y}}_i) \right\} \right]f(\bf y)=0,
\label{eq:appC_EOM}
\end{equation}
where $M$ is the four-dimensional KK mass.
Using the KK expansion
\begin{equation}
f({\bf{y}})= \sum_{l_1,l_2\in\mathbb Z} f_{l_1,l_2} e^{i(\lambda_5x^5+\lambda_6x^6)},
\label{eq:appC_expansion}
\end{equation}
with
\begin{equation}
\lambda_5=\frac{l_1+\beta_1}{R}, \qquad
\lambda_6=\frac{l_2+\beta_2}{R},
\label{eq:appC_lambda}
\end{equation}
the dimensionless shifts are given by
\begin{equation}
\beta_1=\frac{q\alpha_1+\varepsilon}{2}, \qquad
\beta_2=\frac{q\alpha_2+\varepsilon}{2}.
\end{equation}
Here, $q$ is the Wilson-line charge, $\alpha_{1,2}$ are the Wilson-line parameters, and $\varepsilon=0,1$ specifies periodic and antiperiodic boundary conditions, respectively.

Substituting the KK expansion into Eq.~(\ref{eq:appC_EOM}), we obtain
\begin{align}
(\lambda_5^2+\lambda_6^2-M^2)f_{l_1,l_2}  
={M^2}\sum_{i=0}^{3} c_i f({\bf{y}}_i)e^{-i {\bf{\lambda}}\cdot {\bf{y}}_i},
\label{eq:appC_Fourier}
\end{align}
where ${\bf{\lambda}}=(\lambda_5,\lambda_6)$.
For $\lambda_5^2+\lambda_6^2 \neq M^2$, the values of the wave function at the fixed points satisfy
\begin{equation}
f({\bf{y}}_i)=
M^2\sum_{j=0}^{3} c_j \mathcal{G}_{ij}f({\bf{y}}_j),
\label{eq:appC_fixedpoint}
\end{equation}
where
\begin{equation}
\mathcal{G}_{ij}(M)=
\sum_{l_1,l_2\in\mathbb Z}
\frac{e^{i{\bf{\lambda}}\cdot({\bf{y}}_i-{\bf{y}}_j)}}{E^2_{l_1,l_2}-M^2},\qquad E^2_{l_1,l_2}=\lambda_5^2+\lambda_6^2.
\label{eq:appC_Sij}
\end{equation}
A nontrivial solution of Eq.~(\ref{eq:appC_fixedpoint}) requires
\begin{equation}
\det\left[ \delta_{ij}-M^2c_j\mathcal{G}_{ij}(M)\right]=0.
\label{eq:appC_general}
\end{equation}
This is the general secular equation for localized kinetic terms at the fixed points.

We next consider the effect of the BLKTs on the lowest KK mode in the limit of small Wilson-line phases.
For periodic boundary conditions, the lowest eigenvalue is
\begin{equation}
M_0^2=\frac{\beta_1^2+\beta_2^2}{R^2}, \qquad |\beta_1|,|\beta_2|\ll1.
\end{equation}
In this limit, the lowest KK mode gives the dominant contribution to $\mathcal{G}_{ij}(M)$:
\begin{equation}
\mathcal{G}_{ij}(M)\simeq
\frac{e^{i{\bf{\lambda}}_0\cdot({\bf{y}}_i-{\bf{y}}_j)}}{E_{0,0}^2-M^2},
\qquad
{\bf{\lambda}}_0=\frac{1}{R}(\beta_1,\beta_2).
\end{equation}
The corresponding matrix has rank one.
The general KK mass equation reduces to
\begin{equation}
1-\frac{M^2}{E_{0,0}^2-M^2} \sum_{i=0}^{3}c_i=0.
\end{equation}
Therefore, the lowest KK mass is approximately given by
\begin{equation}
M^2\simeq
\frac{\beta_1^2+\beta_2^2}{R^2\left(1+\sum_{i=0}^{3}c_i\right)}.
\label{eq:appC_lowest}
\end{equation}
Thus, at leading order in the small-phase expansion, the lowest KK mass depends only on the sum of the BLKT coefficients, while the dependence on their locations appears at higher orders.

Since the leading-order correction to the lowest KK mass depends only on the sum of the BLKT coefficients, we consider a single BLKT at the origin for simplicity: $c_0=c$ and $c_1=c_2=c_3=0$.
The KK mass equation then reduces to
\begin{equation}
1-cx^2{\cal{S}}(x;\beta_1,\beta_2)=0,
\label{eq:appC_single}
\end{equation}
where $x=MR$ and
\begin{equation}
{\cal{S}}(x;\beta_1,\beta_2)=
\sum_{l_1,l_2\in\mathbb Z} \frac{1}{E^2_{l_1,l_2}-x^2}.
\label{eq:appC_sum}
\end{equation}

The two-dimensional KK sum is logarithmically divergent. We perform the sum over $l_1$ analytically using
\begin{equation}
\sum_{n=-\infty}^{\infty}
\frac{1}{(n+a)^2-z^2}
= \frac{\pi\sin(2\pi z)} {z[\cos(2\pi z)-\cos(2\pi a)]}.
\label{eq:appC_identity}
\end{equation}
Defining
\begin{equation}
z_l=\sqrt{x^2-(l+\beta_2)^2},
\end{equation}
we obtain
\begin{equation}
\mathcal S(x)= \sum_{l=-\infty}^{\infty}F_l(x), \qquad 
F_l(x)= \frac{\pi\sin(2\pi z_l)} {z_l[\cos(2\pi z_l)-\cos(2\pi\beta_1)]}.
\label{eq:appC_S1D}
\end{equation}
For $x^2<(l+\beta_2)^2$, setting
$\kappa_l=\sqrt{(l+\beta_2)^2-x^2}$
gives the equivalent real expression
\begin{equation}
F_l(x)= \frac{\pi\sinh(2\pi\kappa_l)} {\kappa_l[\cosh(2\pi\kappa_l)-\cos(2\pi\beta_1)]}.
\end{equation}
For large $|l|$, the summand behaves as
\begin{equation}
F_l(x)= \frac{\pi}{|l+\beta_2|} +O(|l|^{-3}),
\end{equation}
up to exponentially suppressed terms.
Hence, the partial sum diverges as $2\pi\log N$ for $N\to\infty$.
To isolate the divergent contribution, we introduce the digamma function
\begin{equation}
\psi(z)=\frac{d}{dz}\log\Gamma(z),
\end{equation}
which satisfies
\begin{equation}
\sum_{l=1}^{N}\frac{1}{l\pm a_2} = \psi(N+1\pm a_2)-\psi(1\pm a_2).
\end{equation}
Using the asymptotic behavior at large $N$
\begin{equation}
\psi(N+1\pm a_2)=\log N+O(N^{-1}),
\end{equation}
we define the regularized sum by
\begin{align}
\mathcal S_N^{\rm reg}(x)
=\sum_{l=-N}^{N}F_l(x) -\pi\left[ \psi(N+1+\beta_2) +\psi(N+1-\beta_2) \right].
\label{eq:appC_digamma}
\end{align}
The subtraction removes the logarithmic divergence and specifies the finite part of the KK sum.
The regularized mass equation is
\begin{equation}
1-cx^2\mathcal S_N^{\rm reg}(x)=0.
\label{eq:appC_reg_mass}
\end{equation}
The same subtraction prescription is used throughout the numerical analysis.

Finally, we discuss the degeneracy of the excited KK spectrum.
For a finite KK cutoff, we define
\begin{equation}
\zeta_N(x)=
\prod_{l_1,l_2} (E^2_{l_1,l_2}-x^2).
\label{eq:appC_xi}
\end{equation}
Its logarithmic derivative satisfies
\begin{equation}
\frac{\zeta_N'(x)}{\zeta_N(x)}
=-2x\sum_{l_1,l_2} \frac{1}{E^2_{l_1,l_2}-x^2}.
\label{eq:appC_logderivative}
\end{equation}
Therefore, the unregularized finite-cutoff equation is equivalent to
\begin{equation}
\zeta_N(x)+\frac{cx}{2}\zeta_N'(x)=0.
\label{eq:appC_characteristic}
\end{equation}

Suppose that an unperturbed eigenvalue $E_{l}>0$ has degeneracy $d_l$.
The characteristic polynomial can be factorized as
\begin{equation}
\zeta_N(x)=(E^2_l-x^2)^{d_l}Q_l(x),
\qquad Q_l(E_l)\neq0.
\end{equation}
Substituting this into
Eq.~(\ref{eq:appC_characteristic}), we find
\begin{align}
&\zeta_N(x)+\frac{cx}{2}\xi_N'(x)
\nonumber\\
&=(E^2_l-x^2)^{d_l-1}
\Big[
(E^2_l-x^2)Q_l(x)
-cd_lx^2Q_l(x)+\frac{cx}{2}(E^2_l-x^2)Q_l'(x)
\Big].
\label{eq:appC_factor}
\end{align}
Thus, the original eigenvalue remains a root with multiplicity $d_l-1$.
Only one linear combination of the $d_l$ degenerate modes is shifted by the single BLKT.

This conclusion also holds with the digamma subtraction.
For a finite cutoff, defining
\begin{equation}
D_N=\pi\left[
\psi(N+1+\beta_2)
+\psi(N+1-\beta_2)
\right],
\end{equation}
the regularized characteristic equation can be expressed as
\begin{equation}
\left(1+cx^2D_N\right)\xi_N(x)
+\frac{cx}{2}\xi_N'(x)=0.
\label{eq:appC_reg_characteristic}
\end{equation}
The additional subtraction term does not change the multiplicity $d_l-1$ of the unshifted positive eigenvalues.

We next investigate the location of the shifted KK eigenvalues. For a finite KK cutoff,
we define
\begin{equation}
F(x,c)=1-cx^2
\sum_l\frac{1}{E^2_l-x^2},
\label{eq:appC_root_function}
\end{equation}
where $E^2_l$ are the unperturbed dimensionless KK mass squared eigenvalues.

Differentiating with respect to $x$, we obtain
\begin{equation}
\frac{\partial F}{\partial x}
=-2cx\sum_l
\frac{E^2_l}{(E^2_l-x^2)^2}.
\label{eq:appC_root_derivative}
\end{equation}
For $c>0$, this derivative is negative.
Therefore, $F(x,c)$ is strictly decreasing between consecutive positive poles.

Let the distinct positive poles be ordered as
\begin{equation}
0<E_1<E_2<\cdots<E_s.
\end{equation}
For consecutive poles, we find
\begin{equation}
\lim_{x\to{E_k}^{+}}F(x,c)=+\infty,
\qquad
\lim_{x\to{E_{k+1}}^{-}}F(x,c)=-\infty.
\end{equation}
Consequently, there exists exactly one shifted eigenvalue $x_k$ satisfying
\begin{equation}
{E_k}<x_k<{E_{k+1}}.
\label{eq:appC_interlacing}
\end{equation}
If the lowest unperturbed eigenvalue is positive, an additional shifted eigenvalue lies below the lowest pole.

Furthermore, the dependence of the shifted eigenvalue on the BLKT coefficient is obtained by implicit differentiation:
\begin{equation}
\frac{dx_k}{dc}
=-\frac{\partial F/\partial c}
{\partial F/\partial x}.
\end{equation}
Using $F(x_k,c)=0$, we obtain
\begin{equation}
\frac{dx_k}{dc}
=
-\qty(
2c^2x_k
\displaystyle\sum_l
\frac{E^2_l}{(E^2_l-x_k^2)^2}
)^{-1}<0.
\label{eq:appC_mass_monotonic}
\end{equation}
Thus, each shifted KK mass decreases monotonically as the positive BLKT coefficient $c$ increases.

These properties provide a useful basis for identifying the KK mass eigenvalues in numerical calculations.


\end{document}